\documentclass[ reprint, amsmath,amssymb, aps, onecolumn,11pt,tightenlines]{revtex4-2}
\usepackage{graphicx}
\usepackage{floatrow}
\usepackage{epstopdf, epsfig}
\usepackage{upgreek}
\usepackage{subfigure}
\usepackage{amsmath}
\usepackage{amssymb}
\usepackage{mathrsfs}  
\usepackage{nomencl}
\usepackage[usenames,dvipsnames]{xcolor}
\usepackage{tikz}
\usetikzlibrary{shapes}
\usepackage{MnSymbol}
\usepackage[normalem]{ulem}
\usepackage[ruled,vlined]{algorithm2e}
\usepackage{wrapfig}
\usepackage{colortbl}
 \usepackage{hyperref}
\hypersetup{
	colorlinks=true, 
	linktoc=all,     
	linkcolor=blue,  
	citecolor=blue,
	urlcolor = blue
}

\DeclareRobustCommand\sampleline[1]{%
  \tikz\draw[#1] (0,0) (0,\the\dimexpr\fontdimen22\textfont2\relax)
  -- (2em,\the\dimexpr\fontdimen22\textfont2\relax);%
}

\def\m1line{\vrule width3pt height2.5pt depth -2pt}

\def\bdot{\raise.2em\hbox to .15em{.}}

\usepackage{todonotes}
\def\outline#1{\textcolor{Gray}{}}
	\long\def\comment#1{}

\allowdisplaybreaks[1]

\begin{document}

\title{From limited observations to the state of turbulence: \\
Fundamental difficulties of flow reconstruction}
\author{Tamer A.~Zaki}
\thanks{corresponding author, email: \url{t.zaki@jhu.edu}}
\affiliation{\small Department of Mechanical Engineering, Johns Hopkins University, Baltimore, MD 21218, USA}
\author{Mengze Wang}
\affiliation{\small Department of Mechanical Engineering, Johns Hopkins University, Baltimore, MD 21218, USA}


\begin{abstract}

Numerical simulations of turbulence provide non-intrusive access to all the resolved scales and any quantity of interest, although they often invoke idealizations and assumptions that can compromise realism. In contrast, experimental measurements probe the true flow with lesser idealizations, but they continually contend with spatio-temporal sensor resolution. Assimilating observations directly in simulations can combine the benefits of both approaches and mitigate their respective deficiencies. The problem is expressed in variational form, where we seek the flow field that satisfies the Navier-Stokes equations and minimizes a cost function defined in terms of the deviation of the computational predictions and available observations. In this framework, measurements are no longer a mere record of the instantaneous, local quantity, but rather an encoding of the antecedent flow events that we aim to decode using the governing equations. Chaos plays a central role in obfuscating the interpretation of the data: observations that are infinitesimally close may be due to entirely different earlier conditions. 
We examine a number of state estimation problems: In circular Couette flow, starting from observations of the wall stress, we accurately reconstruct the wavy Taylor vortices that interact nonlinear to maintain a saturated state. 
Through a discussion of transition to chaos in a Lorenz system, we highlight the challenge of navigating the landscape of the cost function, and how the landscape can favorably be modified by sensor weighting and placement.  In turbulent channel flow, the Taylor microscale and Lyapunov timescale place restrictions on the resolution of observations for which we can accurately reconstruct all the missing scales.  
The notion of domain of dependence of an observation is introduced and related to the Hessian of the cost function. For measurements of wall shear stress, the eigenspectrum of the Hessian demonstrates the sensitivity of short-time observations to the fine near-wall turbulent scales and to the large scales only in the outer flow.  At long times, backward chaos obfuscates the interpretation of the data: observations that are infinitesimally close may be traced back to entirely different earlier flow states\textemdash a dual to the more common butterfly effect for forward trajectories.  

\end{abstract}

\maketitle

\section{Introduction}

\noindent

\subsection{Why data assimilation?}

Flow simulations are continually pushing computational boundaries to establish new frontiers, for example turbulence at progressively higher Reynolds numbers \citep{Yeung2018dns,Lee2015dns} and phenomena that take place at hypersonic speeds \citep{marxen2014direct}. 
Simulations also attempt to probe ever more complex configurations that involve multiphase, rheological, physiological effects \citep{esteghamatian_2021,lee_2017,Seo2013cardio,nicolaou2013}.
Despite impressive advancements, even at the highest levels of simulation fidelity some modeling assumptions invariably remain, including treatment of boundary conditions such as periodic and truncated simulation domains \citep{Yang2016inflow,Wu2017inflow} and initial conditions when multiple flow states exist for the same set of parameters \citep{Lohse2014TC}.  At extreme conditions, models are adopted to reduce the computational cost or to account for unknown physics. Choices of model parameters, for example in sub-grid stress and wall models in large-eddy simulations of high Reynolds number or in non-equilibrium chemical models at high speeds, have direct impact on the accuracy and robustness of simulations.

Experiments on the other hand provide a level of realism that is difficult to reproduce computationally, and can access extreme flow regimes, including very high Reynolds numbers that remain, to date, intractable computationally \citep{Smits2011review}.
Experiments must, however, contend with the difficulty of performing non-intrusive measurements and the limitations of sensor resolution.  In some instances, it is not possible to directly measure a quantity of interest and it can only be inferred or computed from other measurements.  For example, while we may be interested in the pressure field in a turbulent flow, the experimental measurements may only provide successive images of particle tracers that must be interpreted and processed to compute the velocity field and subsequently the associated pressure \citep{Katz2019pressure}.

Data assimilation combines the respective strengths of both experiments and simulations, and mitigates their deficiencies \citep{Stuart2015,mons2016reconstruction}:  
By infusing experimental measurements in simulations, we enhance the fidelity of the computations and mitigate epistemic uncertainties.  
At the same time, the numerical prediction are not only consistent with the observations, 
they also provide non-intrusive access to the full flow state at much higher resolution than the original data \citep{mengze2021}.  
The benefit of solving these inverse problems, from observations to flow state, extends beyond simply enhancing the resolution of scarce data. 
Inverse problems provide a unique perspective for the interpretation of observations. 
A sensor signal is no longer a mere record of a measured quantity at a point in space and an instance in time, but rather an encoding of the antecedent flow events that led to the measurement \citep{wang_hasegawa_zaki_2019}.  
By decoding the observations we reveal the hidden flow physics that is unavailable from forward analysis, including the domain of dependence of an observation site, the critical data resolution for an accurate state estimation, and the robustness of reconstructed flow events to observation noise. 
This perspective also motivates a new notion of optimal layout and weighting of measurements: 
Instead of concentrating measurements near a region of interest based on knowledge of the forward problem, sensor placements can be optimized in order to provide the most accurate interpretation, or reconstruction, of the entire flow field using data assimilation \citep{mons2019kriging,buchta2021envar}.

\subsection{Choice of approach}

In choosing a data-assimilation approach for the interpretation, or augmentation, of observations, it is important to differentiate some of the existing classes of techniques. 
One distinction is among methods that actually satisfy the governing flow equations versus those that approximate the velocity field using a statistical or reduced-order, perhaps machine-learned, model \citep{encinar2019logarithmic,Chandramouli2020,Raissi2019PINN,Du2021EDNN}.
Here our focus will be on predictions that satisfy the full Navier-Stokes equations and that best reproduce observations.
This choice is desirable if one is interested in the fundamental flow dynamics, as we are here, or in forecasting beyond the observation time horizon.

Another classification is based on how observations are assimilated into numerical simulations. 
Filters use the latest observations to update the state \citep{Evensen1994EnKF,Suzuki2012}, while smoothers attempt to identify the state whose evolution reproduces all the observations within the entire assimilation horizon \citep{Dimet1986_4dvar,mons2016reconstruction}.  
Again, since our interest is in the evolution of the flow state, or the dynamics, we will focus on smoothers. 
In this context, we define a cost function which is the difference between model predictions and observations, and we proceed to identify the flow state that (i) satisfies the governing Navier-Stokes equations and (ii) minimizes the deviation from observations. 
The minimization procedure requires the gradient of the cost function with respect to the initial flow state;
this gradient also quantifies the sensitivity of observations to initial flow field.

Two smoothing techniques have been adopted successfully in turbulent flows: adjoint- and ensemble-variational methods.
Adjoint methods are the most efficient in terms of evaluating the gradient \citep{mengze2019discrete}.
The associated computational cost is always on the order of twice that of calculating the cost function, regardless of the size of the control vector.
However, for chaotic systems the accuracy of the gradient deteriorates for long integration time horizons \citep{chandramoorthy2019feasibility,eyink2004ruelle}. 
Therefore, the implementation of adjoint methods is restricted to short assimilation windows that can be concatenated to span long periods of observations \citep{Fisher2012,Chandramouli2020}. 
Ensemble methods do not require an adjoint model \citep{mons2019kriging}.  
They use an ensemble of estimates of the state and advance them in time, collect model predictions and compare them to observations.  The local gradient of the cost function with respect to the state can be approximated at the location of the mean of the ensemble using a quadratic approximation, and used to update the estimated state. 
These methods are suitable for long horizons, for example when using statistical observations.  
The number of ensemble members, however, increases with the size of the control vector and they are not necessarily as robust as adjoint techniques.
Some methods combine the idea of ensemble and adjoint to evaluate the sensitivity of statistical observations to model parameters \citep{eyink2004ruelle}.
Here we adopt adjoint-variational method, because we are interested in fundamental questions regarding the domain of dependence of observations, the minimum observation resolution that enable reconstruction of the entire turbulent field, and the robustness of reconstruction to observation noise.

\vspace*{6pt}
In the next section, we introduce the adjoint-variational approach briefly.  
We proceed to apply the method to the interpretation of measurements in circular Couette flow.  
The configuration is designed such that many saturated states are possible, and we attempt  to identify the particular one that generated the wall stresses.  
In this case, the dynamics are not chaotic. 
We proceed to discuss what to expect in case of transition to turbulence, and briefly remark on some recent results in that domain.  
We then turn to canonical turbulent channel flow, explore the critical data resolutions that enable an accurate turbulence reconstruction, and discuss the notion of domain of dependence of measurements.
Concluding remarks are provided in the final section.

\section{Adjoint variational data assimilation}
\label{sec:4DVar}

Since our interest is in a high-fidelity interpretation of flow observations, throughout we will adopt the Navier-Stokes equations to predict the evolution of the flow and will resolve all the flow scale using direct numerical simulations.  This choice eliminates any turbulence modeling assumptions. Our flow thus satisfies the incompressibility constraint and momentum equations, 
\begin{eqnarray}
	\label{eq:cont_div}
       \nabla \cdot \boldsymbol{u} &=& 0  \\
	\label{eq:cont_mom}
       \frac{\partial \boldsymbol{u} }{\partial t} +  \boldsymbol u \cdot \nabla \boldsymbol u &=& - \nabla p + \frac{1}{Re} \nabla^2 \boldsymbol{u}. 
\end{eqnarray}
These equations are also referred to as the forward model in adjoint-variational method, and the velocity $\boldsymbol u$ and pressure $p$ as the forward state variables.

In lieu of experimental measurements, we perform independent simulations from a reference initial condition $\boldsymbol u^R_0$ and extract surrogate observations,
\begin{equation}
    \label{eq:obsevation}
    \boldsymbol m_n = \mathcal M(\boldsymbol u^R _n).
\end{equation}
The reference initial state $\boldsymbol u^R_0$ then becomes the hidden truth that we aim to discover from its observations $\boldsymbol m_n$ alone.
The objective of adjoint-variational data assimilation is to identify an estimated initial condition $\boldsymbol u_0$ that reproduces the observations, by minimizing a cost function that is proportional to the difference between the observations and their estimation,
\begin{equation}
    \label{eq:cost}
	J(\boldsymbol u_0) = \sum_{n=0}^N \frac 12 \|\boldsymbol m_n - \mathcal M(\boldsymbol u_n) \|^2_O. 
\end{equation}
The norm $\| \cdot \|_O$ represents evaluation over the observation space, and $\boldsymbol u_n$ is the Navier-Stokes evolution of the estimated initial condition.   
The estimated initial condition is sought iteratively using a gradient-based minimization of the cost function.  

In order to evaluate the  gradient of $J$, we introduce the Lagrangian 
\begin{equation}
	\label{eq:cont_lagrangian}
       L = J - \left[\boldsymbol{u}^{\dag},  \frac{\partial \boldsymbol{u} }{\partial t} + \boldsymbol u \cdot \nabla \boldsymbol u + \nabla p - \frac{1}{Re} \nabla^2 \boldsymbol{u}\right] - \left[p^{\dag}, \nabla \cdot \boldsymbol{u} \right], 
\end{equation}
where the multipliers $\boldsymbol u^{\dag}$ and $p^{\dag}$ are the adjoint velocity and adjoint pressure, respectively.
The inner product is an integration over the spatial domain and the time horizon, 
\begin{equation}
	\label{eq:cont_inner}
     \left[\boldsymbol f, \boldsymbol g \right] = \int_{0}^{T} \int_V \boldsymbol f ^{\top} \boldsymbol g \ d^3 \boldsymbol x dt, 
\end{equation}
where $t=0$ and $t=T$ represent the start and end of the assimilation window. 
First-order optimality requires that the derivatives of $L$ with respect to both the adjoint and forward variables be zero. 
The former condition yields the Navier-Stokes equations (\ref{eq:cont_div}-\ref{eq:cont_mom}) which are enforced by simulating the forward evolution of the estimated field.
The latter condition, namely that the derivatives of $L$ with respect to $\boldsymbol u_n$ vanishes,
yields the adjoint equations, 
\begin{eqnarray}
	\label{eq:cont_adj_div}
       \nabla \cdot \boldsymbol u^{\dag} &=& - \frac{\partial J}{\partial p} \\
	\label{eq:cont_adj_mom}
         \frac{\partial \boldsymbol u^{\dag}}{\partial \tau} - \boldsymbol u \cdot \nabla \boldsymbol u^{\dag} + (\nabla \boldsymbol u) \cdot \boldsymbol u^{\dag} &=& \nabla p^{\dag} + \frac{1}{Re} \nabla^2 \boldsymbol u^{\dag} + \frac{\partial J}{\partial \boldsymbol u}, 
\end{eqnarray}
where $\tau \equiv T - t$ is the reverse time. 
The adjoint system is driven by the source term, which is derived analytically from the cost function (\ref{eq:cost}) and determined by the mismatch between the observations and their estimation, $\mathcal E_n = \mathcal M(\boldsymbol u_n) - \boldsymbol m_n$.

\begin{figure}
    \centering
    \includegraphics[width = 0.9\textwidth]{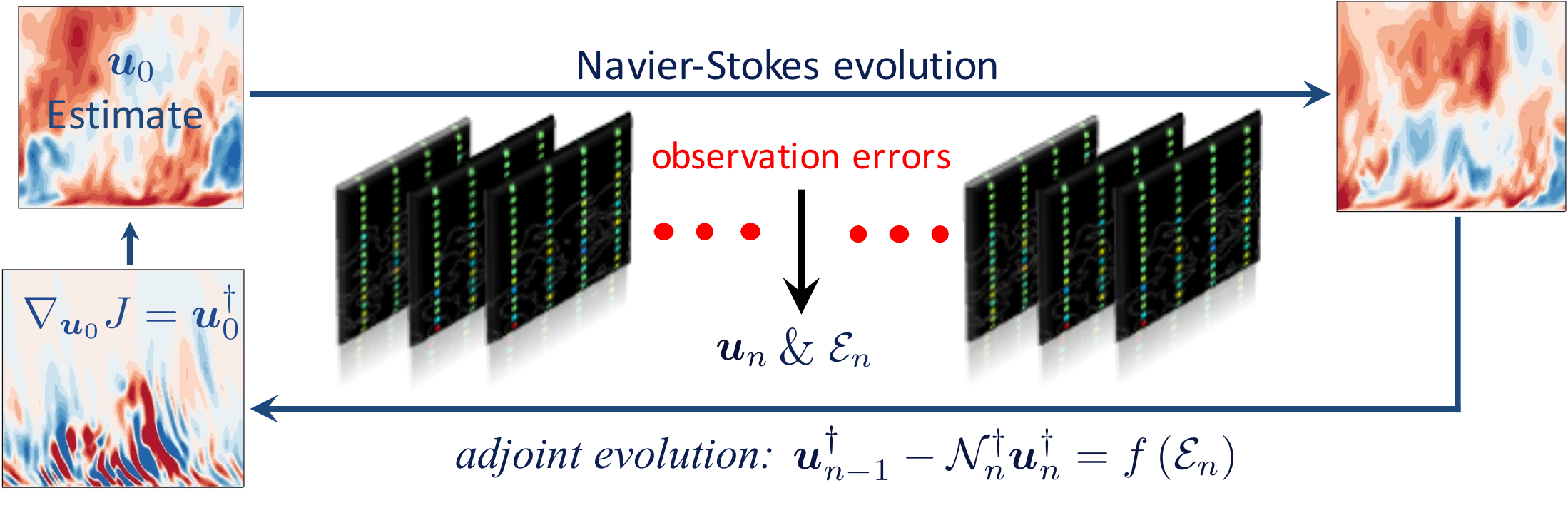}
    \caption{
    Schematic of adjoint-variational data assimilation. Starting from an estimate of the initial condition $\boldsymbol u_0$, we compute the Navier-Stokes evolution. The forward fields $\boldsymbol u_n$ and the deviation from observations $\mathcal E_n$ are stored.
    The adjoint equations are evolved from the final to the initial time, driven by observations errors.
    The initial adjoint field $\boldsymbol u_0^{\dag}$ is the gradient of the cost function, and is used to update $\boldsymbol u_0$. The forward-adjoint loop is repeated until convergence.
    }
    \label{fig:4DVar}
\end{figure}

The gradient of the cost function is determined by the adjoint field at the initial time $t=0$ ($\tau=T$),
\begin{equation}
    \label{eq:grad}
    \nabla_{\boldsymbol u_0} J = \frac{\partial L}{\partial \boldsymbol u_0} = \boldsymbol u^{\dag}_0.
\end{equation}
Therefore, the adjoint field $\boldsymbol u^{\dag}_0$ is the sensitivity of the cost function to initial conditions.
This sensitivity can then be used in gradient-based optimization algorithms to minimize $J$.
In the examples presented in this paper, we adopt the limited-memory Broyden-Fletcher-Goldfarb-Shanno (L-BFGS) method \cite{LBFGS}.
The above presentation was in terms of the continuous form of the equations.  Discretization of the continuous forward and adjoint systems may affect the accuracy of evaluating the gradient of the cost function.  To obtain the exact gradient, we adopted the discrete adjoint approach where we first discretize the forward Navier-Stokes equations and then derive the discrete adjoint system.

The adjoint-variational data assimilation procedure is presented schematically in figure \ref{fig:4DVar}.
Starting from an estimate of the initial flow state $\boldsymbol u_0$, the Navier-Stokes equations are evolved over the assimilation time horizon.  During the forward evolution, both the velocity fields and the discrepancies between the observations and their estimates are stored.  At the end of the assimilation horizon, the adjoint equations are advanced backward in time, forced by the errors in the observations. The resulting gradient of the cost function at the end of adjoint marching is used in the L-BFGS algorithm to update the estimated initial condition.  This procedure is then repeated for a prescribed number of iterations or until a convergence criterion is reached.

The utility of data assimilation is far reaching, beyond the important application of augmenting experimental measurements.  These techniques enable us to pose new questions. In turbulence, for instance, we can progressively increase the sparsity of observations and ask what is the largest spatio-temporal spacing that still enables full reconstruction of the missing scales, and why?  
The adjoint-variational approach in particular provides a unique perspective for interpretation of the flow physics.  Adjoint fields quantify the sensitivity of observations to the flow state.
When observations are spatio-temporally distributed, comparing their adjoint fields provides guidance for sensor placement and relative weighting, and motivates new strategies to optimize these sensing parameters.
Furthermore, adjoint equations can be derived to evaluate the domain of dependence of an isolated measurement and the Hessian of the cost function at optimality, which yield a unique perspective on the difficulty of estimating the state of turbulence from limited observations.  These ideas will be explored in the remaining sections with the aid of specifically designed examples.

\section{Flow reconstruction from limited measurements}
\label{sec:results}

\subsection{Traveling saturated waves in circular Couette flow}

The first configuration that we will examine is the circular Couette flow between two concentric cylinders, which was explored in detail by  \citet{mengze2019discrete}.  The gap between the inner and outer cylinders is $d = r_o - r_i = 1$, and the ratio of radii is $r_i/r_o=0.714$.  
The outer cylinder is stationary, and the inner-cylinder tangential speed $v_i$ is adopted as the reference velocity scale in the definition of the Reynolds number, $Re_i \equiv v_i d / \nu = 400$. 
The above parameters were selected such that we can demonstrate the method in a flow that involves nonlinearly interacting, unsteady, traveling waves, yet the total energy is constant and the flow is not chaotic.  Another important feature is that for the same Reynolds number, there are multiple admissible flow states and which one is observed depends on the precise experimental geometry and protocol, for example the aspect ratio of the cell, the initial condition and the transients.  
Simulations may not be able to reproduce all the experimental conditions, which may also include uncertainties.  
However, using observations from the experiments within the saturated flow regime only, i.e.\,beyond any experimental transients, we can adopt data assimilation strategies to reproduce the exact experimental state and the precise evolution of the constituent waves as a function of time.  

The observations that we attempt to interpret are the shear stresses on the cylinder walls as a function of time, which are imprinted by the signature of the nonlinearly interacting, traveling, wavy Taylor rolls.  These observations spanned a time horizon $T=20$ time units, which is more than one rotation of the inner cylinder, and were collected from an independent computation. The associated flow is the hidden truth that we aim to discover.  

Our first estimate of the flow state can simply be the laminar profile between two concentric cylinders.  This state naturally fails to reproduce the observations since it is void of any Taylor vortices. At the end of the first forward simulation, the adjoint model is then forced by the discrepancy in the wall shear stresses from the first guess and available observations.  Upon reaching the initial time, the adjoint field provides the gradient of the cost function with respect to the initial condition, which is used in the L-BFGS algorithm to obtain the update direction of the initial flow state.  This procedure is repeated iteratively until our computational predictions reproduce the available data of the wall stresses to within a prescribed level of accuracy. 

\begin{figure}
    \centering
    \includegraphics[width = \textwidth]{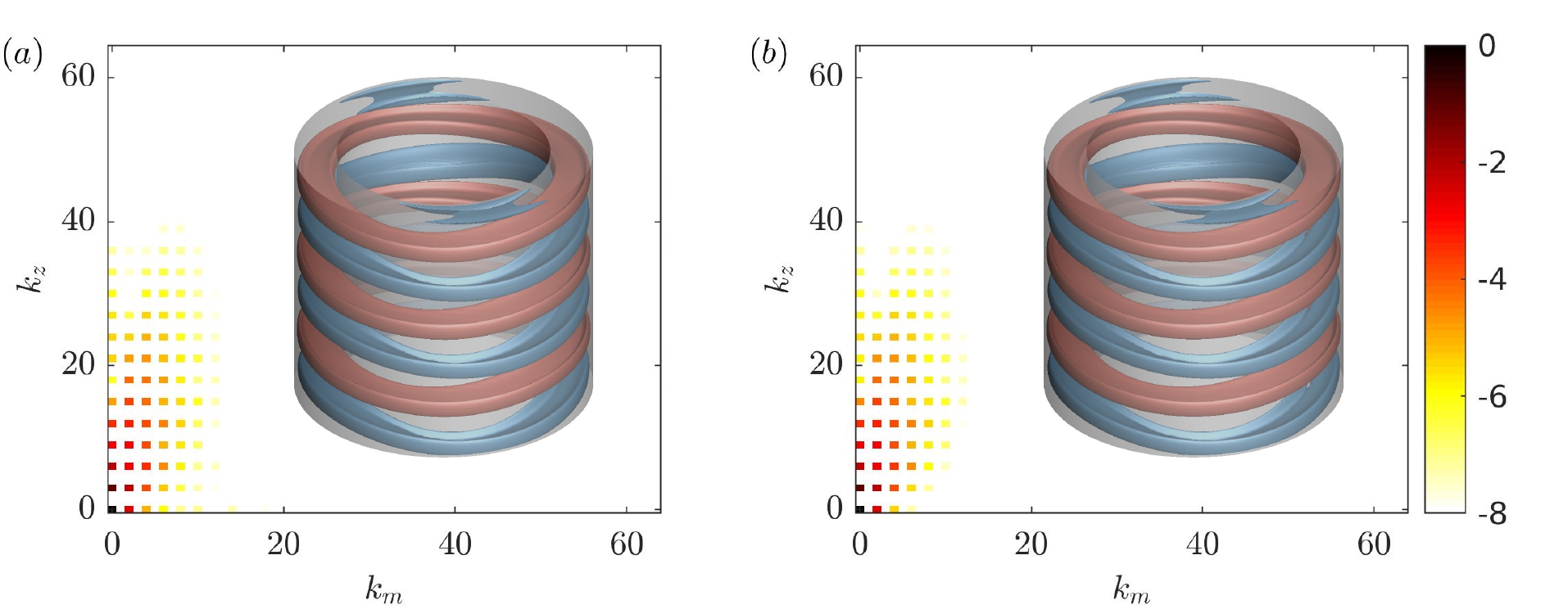}
    \caption{Energy spectra averaged across the gap, $\log_{10} \langle \|\hat {\boldsymbol u} \|^2\rangle_d$, where $\hat {\boldsymbol u}$ is the Fourier transform of the velocity perturbation field $\boldsymbol u - \langle \boldsymbol u^R\rangle$. 
    Insets: isosurfaces of (red) positive and (blue) negative azimuthal velocity fluctuations, $v_{\theta} - \langle v_{\theta}^R\rangle = \{0.14, -0.14\}$.
    ($a$) Adjoint-variational estimation at $t=T$; ($b$) the true state.
    Both spectra are normalized by the maximum value of the true spectrum.
    }
    \label{fig:TC}
\end{figure}

Results reported here were obtained after one hundred forward-adjoint loops in order to ensure accuracy. The resulting decrease in the cost function was three orders of magnitude from its initial value.  The predicted state $\boldsymbol{u}$ at the end of the assimilation time horizon is visualized in figure \ref{fig:TC} where it is compared to the hidden truth $\boldsymbol{u}^R$.  
The iso-surfaces show positive and negative tangential velocity perturbations.
Qualitatively, the estimated flow has the correct number of wavy Taylor rolls and azimuthal phase of the structures.  Quantitatively, the root-mean-squared error $\mathcal E_V=\langle\|\boldsymbol{u}-\boldsymbol{u}^R\|^2\rangle_{V}^{1/2}$, averaged over the volume $V$, is less than $5\%$.  
The spectra of the assimilated and true states are also compared at the final time. This plot  highlights the presence of a large number of axi-symmetric and wavy vortices, traveling around the azimuth and interacting nonlinearly. The agreement in the spectra between the assimilated and true states is evident, in particular for energetic modes that represent dominant flow structures. The scale in the figure is selected to highlight the accuracy of the prediction across eight orders of magnitude.

This first example is by no means trivial as can be inferred form the energy spectra.  The results demonstrate that data assimilation can be adopted to augment limited experimental observations, and mitigate uncertainty in computational parameters.  Specifically, the adjoint approach did not require any knowledge of the startup procedure of the true flow that led to the saturated state when the measurements were collected.  Instead, the forward-adjoint loop relied on the measurements to directly discover the initial condition within the energy saturated state and its time evolution.  It should be noted, however, that this first example was designed to eliminate a challenge, namely estimation of chaotic dynamics.  Specifically, in the present case, once the energy of the system reaches the saturated state, the flow is stable.  What happens when the flow is chaotic?  Before tackling the problem of estimating a fully turbulent state, it is instructive to examine a transitional system where organized and chaotic dynamics are juxtaposed.

\subsection{Transition to chaos}
\label{sec:transition}

Laminar-to-turbulence transition is difficult to predict due to its chaotic nature.  Infinitesimal changes in the system state can lead to significant modification of transition characteristics.  
For example, spontaneously formed turbulent spots in bypass transition of boundary layers \citep{zaki2013streaks,marxen2019turbulence,wupnas2020} and puffs in pipe flows \citep{Reynolds1883,Avila2011pipe} appear sporadically and are difficult to anticipate. 
As the system undergoes transition from an organized to a chaotic state, it is intuitive to expect that early observations are more valuable, or beneficial, when attempting to estimate the initial condition. A cost function ${J}$ defined using sensors in the laminar region will vary smoothly when estimates of the initial state approach the truth, which can promote convergence of the optimization procedure. In contrast, when sensors are in the turbulent region, the cost function oscillates significantly in the vicinity of the measurements since small changes in the initial condition can lead to significantly different observations and, as a result, convergence may be more difficult.  

\begin{figure}
    \centering
    \includegraphics[width = 0.9\textwidth]{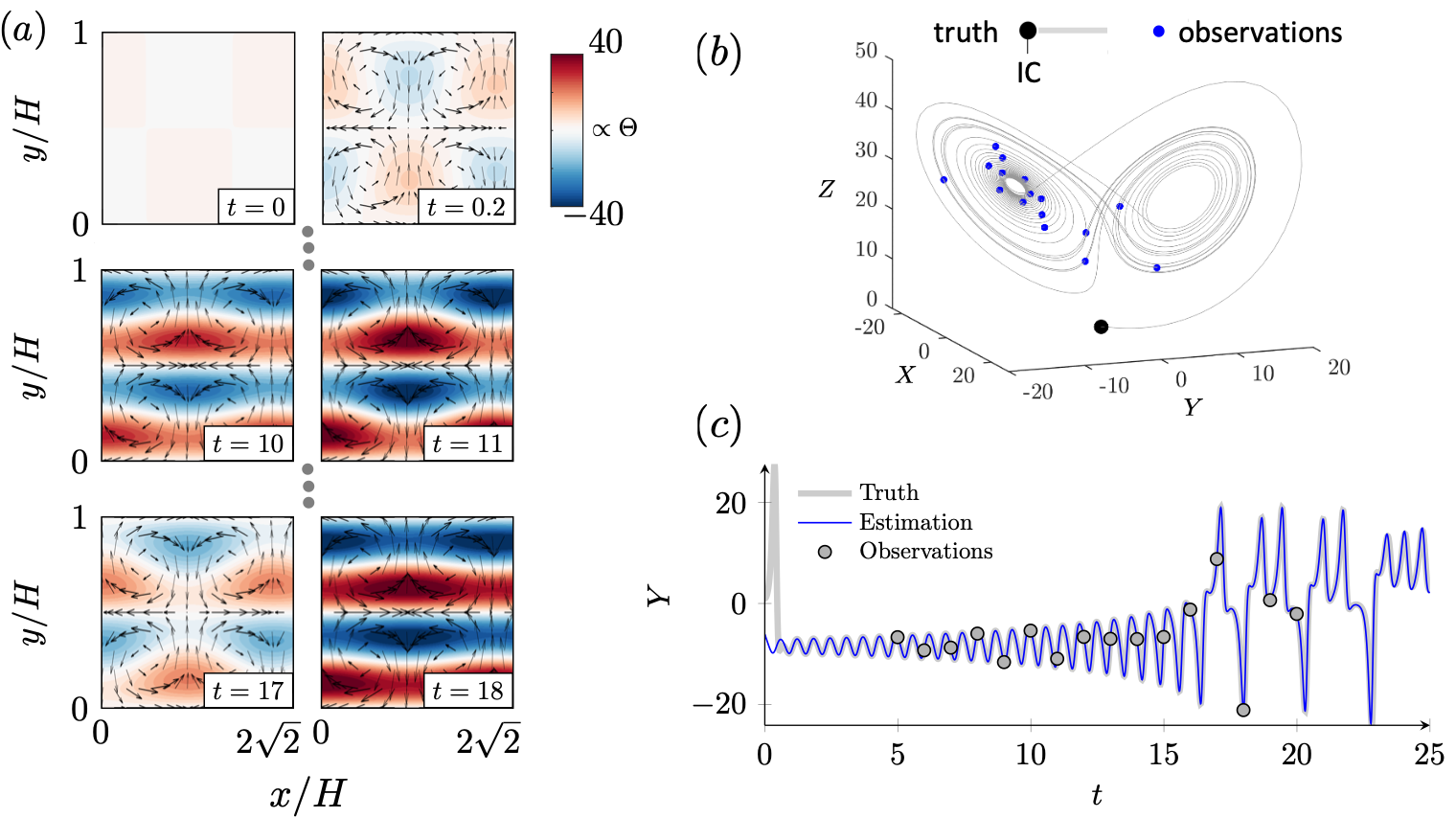}
    \caption{(a) Evolution of the Lorenz system in physical space. Contours are the temperature perturbations and arrows show the velocity field.  (b) State-space evolution of the Fourier coefficients, starting from the true initial condition with marked observations. ($c$) Time evolution of $Y$ from the (gray) true and (blue) estimated initial conditions, with marked observations.  The assimilated trajectory latches onto the truth, reproduces observations, and accurately forecasts the system for some time beyond the observation horizon \citep{buchta2021envar}.}
    \label{fig:lorenz}
\end{figure}

A helpful prototype system to explore the influence of transition and chaos on data assimilation is the Lorenz model of unstable convection in two dimensions \citep{buchta2021envar}.  The setup is periodic in the horizontal direction, and both gravity and a temperature difference between the lower and upper boundaries, $\Delta \Theta=\Theta(0)-\Theta(H)$, act in the vertical direction.   Small initial perturbations amplify in time, and lead to convection cells that oscillate about an equilibrium and suddenly overturn (figure \ref{fig:lorenz}). The process of oscillation and overturning repeats in a chaotic fashion \citep{Lorenz1963}. The early amplification of perturbations is akin to growth of instability waves in wall-bounded flows, and the resulting sporadic overturning of the cells is a surrogate for the chaotic inception of turbulence.

For the Lorenz system, the temperature fluctuations $\Theta$ and velocity potential $\psi$ can be expressed in Fourier space in terms of the horizontal and vertical wavenumbers $k_a = a 2\pi/H$ and $k_y = 2\pi/H$,  
\begin{align}\label{eq:tpsi}
    \Theta(x,y,t)&\propto Y(t) \sqrt{2} \cos\left(\tfrac{1}{2}k_a x \right)\sin\left(\tfrac{1}{2}k_y y \right)-Z(t)\sin\left(k_y y \right)\\\label{eq:velocitypotential}
    \psi(x,y,t)&\propto X(t)\sqrt{2} \sin\left(\tfrac{1}{2}k_a x \right)\sin\left(\tfrac{1}{2}k_y y\right). 
\end{align}
The time-evolution of the Fourier coefficients $\boldsymbol{Q} = [X,Y,Z]^\top$ is given by,
\begin{align}\label{eq:Lorenz}
    \frac{dX}{dt}=\sigma(Y-X),\quad
    \frac{dY}{dt}=-XZ+\rho X -Y,\quad\textrm{and}\quad 
    \frac{dZ}{dt}= XY - \beta Z.
\end{align}
Similar to \citet{Saltzman1962}, we adopt the initial condition $[X_o, Y_o, Z_o]^\top = [0, 1,  0]^{\top}$ which becomes our hidden truth, and we consider $a=\sqrt{1/2}$, $\sigma=10$, $\rho=28$.  The small size of the system enables us to efficiently evaluate and visualize the cost function, and to explore ideas related to sensor placement and weighting \citep[see also appendix by][]{buchta2021envar}.

\begin{figure}
    \centering
        \includegraphics[width = 0.90 \textwidth]{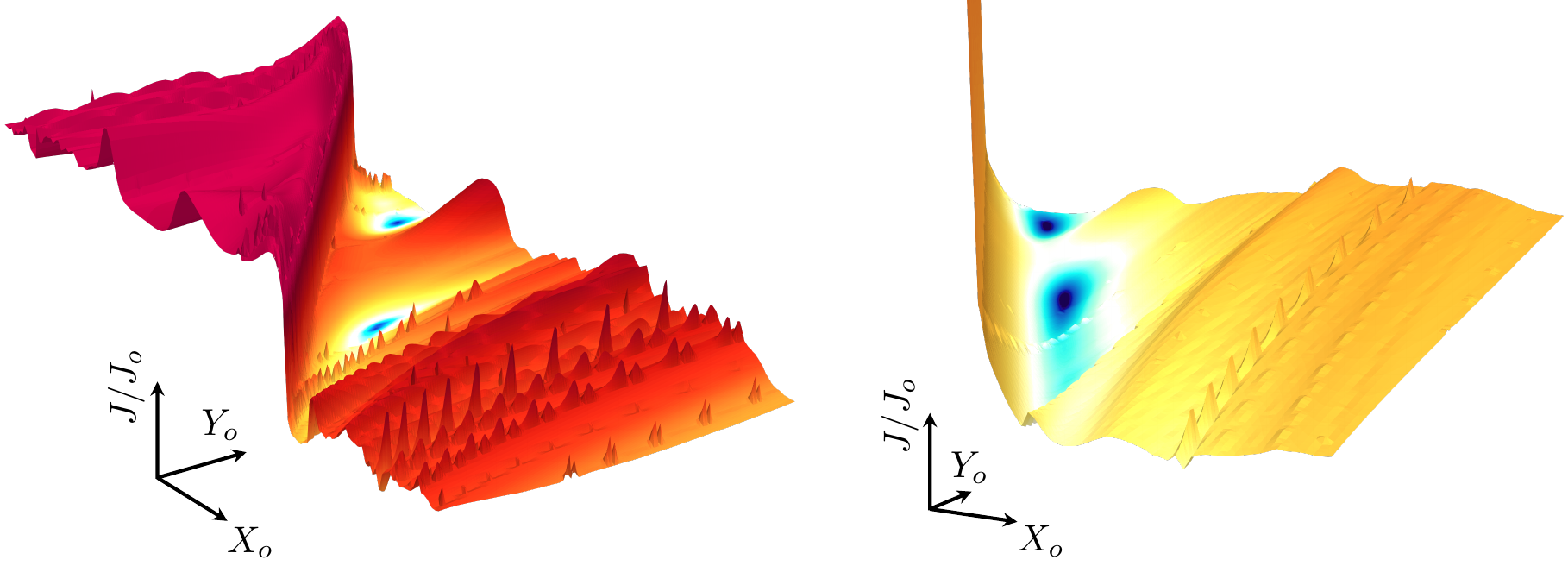}
    \caption{The dependence of the cost function of the Lorenz system on $(X_o, Y_o)$ near the global optimal.  
    Left: The landscape of the cost function when all observations are uniformly weighted.
    Right: The landscape of $J$ when the observations are weighted by a Gaussian that preferentially weights early measurements.}
    \label{fig:cost}
\end{figure}

Consider when observations are available for $Y$ only, at time instances $t_m = \{5, 6, \ldots, 20\}$ which span the early periodic solution and later chaotic behavior (see figure \ref{fig:lorenz}). We attempt to discover the initial condition $[X_o, Y_o, Z_o]^\top$ that reproduces the measurement vector  $\boldsymbol{m}=\left[Y(t_m)\right]^\top$ by minimizing the cost function $J=\frac{1}{2}\|\boldsymbol{m}-\mathcal{M}(Y)\|^2$.  For this small system, we can directly evaluate $J$ and plot its dependence on the control vector near our estimate. Figure \ref{fig:cost} shows that the cost-function landscape is highly oscillatory, laden with local minima and, as a result, difficult to navigate to identify the global minimum.  The idea of preferentially weighting early observation was evaluated by \citet{buchta2021envar}.  
The weighted cost function is $J=\frac{1}{2}\|\boldsymbol W\left(\boldsymbol{m}-\mathcal{M}(Y) \right)\|^2$, where $\boldsymbol W$ is a diagonal matrix with larger elements corresponding to earlier observations.
The resulting cost-function landscape (figure \ref{fig:cost}) is much smoother and amenable to efficient gradient-based optimization. 

Using the preferential weighting of upstream observations, \citep{buchta2021envar} performed an ensemble, rather than adjoint, variational estimation of the initial condition and the results are shown in figure \ref{fig:lorenz}$c$.  While the predicted initial state is visibly displaced relative to the truth, the predicted trajectory quickly latches onto the true one, reproduces all the observations and provides an accurate forecast for the shown duration past the observation horizon.

While the idea of sensor weighting was motivated here by physical arguments, it is based on rigorous mathematical foundation.  
The shape of the cost function depends on the observation operator $\mathcal M$, which determines the measured quantities and their placements, and the measurement weighting $\boldsymbol W$.  
Since at optimality the gradient of $J$ vanishes, we are particularly interested in how the design of observations affects the Hessian of the cost function.  Preferentially weighting early measurements reduces the maximum eigenvalue of the Hessian, thus reducing extreme curvature.  \citet{buchta2021envar} adopted this strategy not only for the Lorenz system above, but also for the much more complex and higher dimensional configuration a transitional, high-speed boundary layer.  They were able to achieve significant improvement in the accuracy of inflow estimation from discrete wall-pressure probes by preferentially weighting sensors upstream of transition.  
It should be cautioned, however, that weighting early observations should not be naively adopted.  In a transitional boundary layer, for instance, a wall-pressure probe should not be placed arbitrarily upstream. An obvious consideration is the potentially low signal-to-noise ratio. Additionally, if placed ahead of the receptivity site or, more subtly, in a region where the sensor does not have sensitivity to the incoming instability wave, the associated observations can not assist in minimizing the cost function. Also note that, in a fully turbulent flow, the cost landscape is mountainous independent of the choice of observed quantities, their spatio-temporal placement and their relative weighting.  In that case,  optimal sensor placement and weighting can target improving the condition number of the Hessian, which has been adopted with success in problems involving prediction of scalar sources in turbulent environments \citep{mons2019kriging,wang_hasegawa_zaki_2019}. We will return to a discussion of the Hessian after we examine the accuracy of adjoint-variational estimation of channel-flow turbulence from limited observations.

\begin{figure}
    \centering
    \includegraphics[width = 0.80 \textwidth]{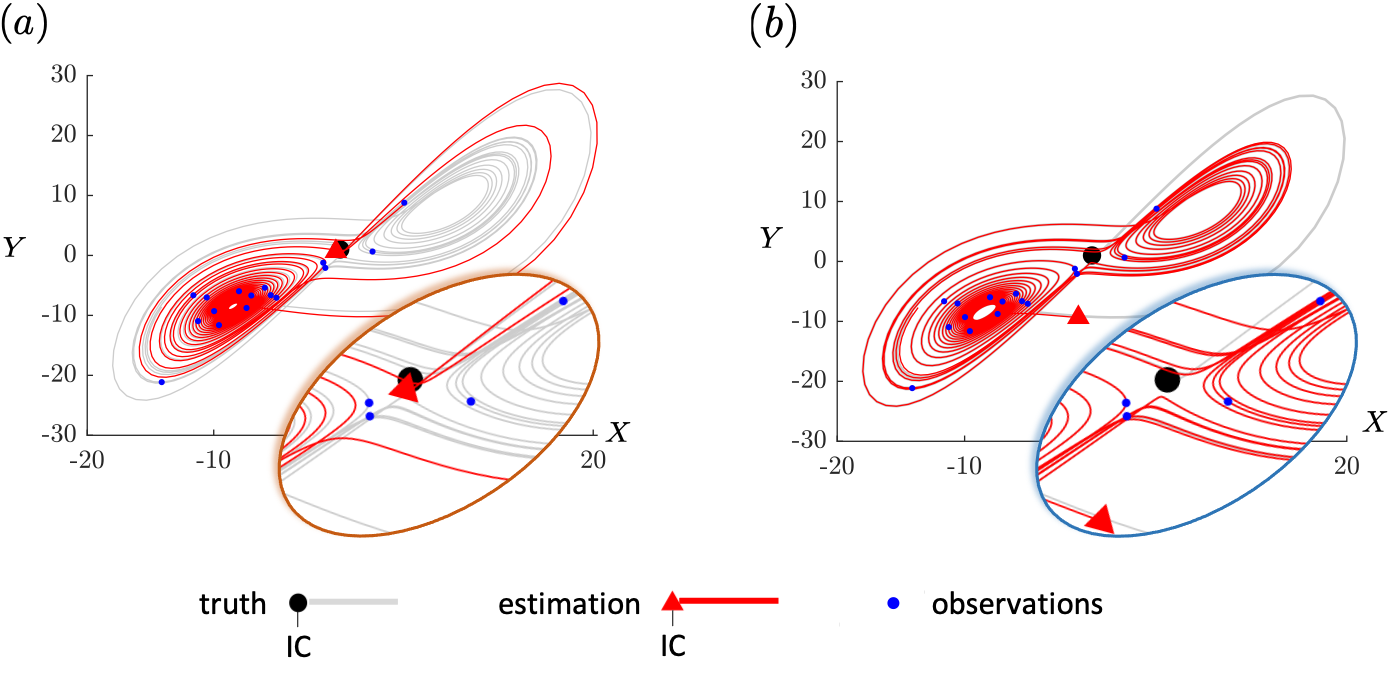}
    \caption{State-space evolution from estimates of the initial Fourier coefficients of the Lorenz system.  ($a$) Divergence of trajectories starting from the true and a slightly perturbed initial condition. ($b$) Trajectory from estimated initial condition.  Despite a large initial separation from the truth, the assimilated trajectory latches onto the truth, reproduces observations, and accurately forecasts the system for some time beyond the observations horizon.}
    \label{fig:butterfly}
\end{figure}

As we turn to the fully turbulent case, we should keep in mind important points that are well captured by the results of the Lorenz system.  
Firstly, due to the intricate landscape and numerous local minima of the cost function (figure \ref{fig:cost}), the first guess, or estimate, of the state in the variational algorithm has an impact on convergence and, hence, the final estimated state.
Therefore, the construction of the first guess should exploit any known characteristics of the initial state.
Absent such knowledge, multiple starting points for the assimilation procedure can be adopted.
Secondly, for chaotic nonlinear systems, there is no guarantee of a unique solution to the inverse problem. The results in figure \ref{fig:lorenz}$c$ show that measurements, and extended portions of the trajectory, can be accurately reproduced from more than one initial condition.
This point can be further clarified by contrasting the ``butterfly effect" and its dual, or ``adjoint butterfly effect", which are illustrated in figure \ref{fig:butterfly}.
In the former case, a small change in the initial condition leads to significant deviation in the forward evolution due to chaos, and the same is of course true of turbulence. In practical terms, a simulation that starts from a perfectly measured state will diverge from the experiment and from other simulations that have as little as different round-off errors.  This behavior can impede our attempt to predict the initial condition that reproduces the measurements, limit the assimilation time horizon and compromise the accuracy of forecasts.  As for the dual butterfly effect, it may be less familiar but is crucially important, especially for the adjoint approach. Infinitesimally close observations can be reproduced by initial conditions that have large separation as we remarked earlier.  In terms of the adjoint, the small deviations in the observations lead to entirely different adjoint evolution, and are traced back to different initial conditions. From a practical standpoint, errors in the observations can thwart accurate estimation of the state.

\subsection{Turbulent channel flow}
\label{sec:channel}

\begin{figure}
    \centering
    \includegraphics[width = \textwidth]{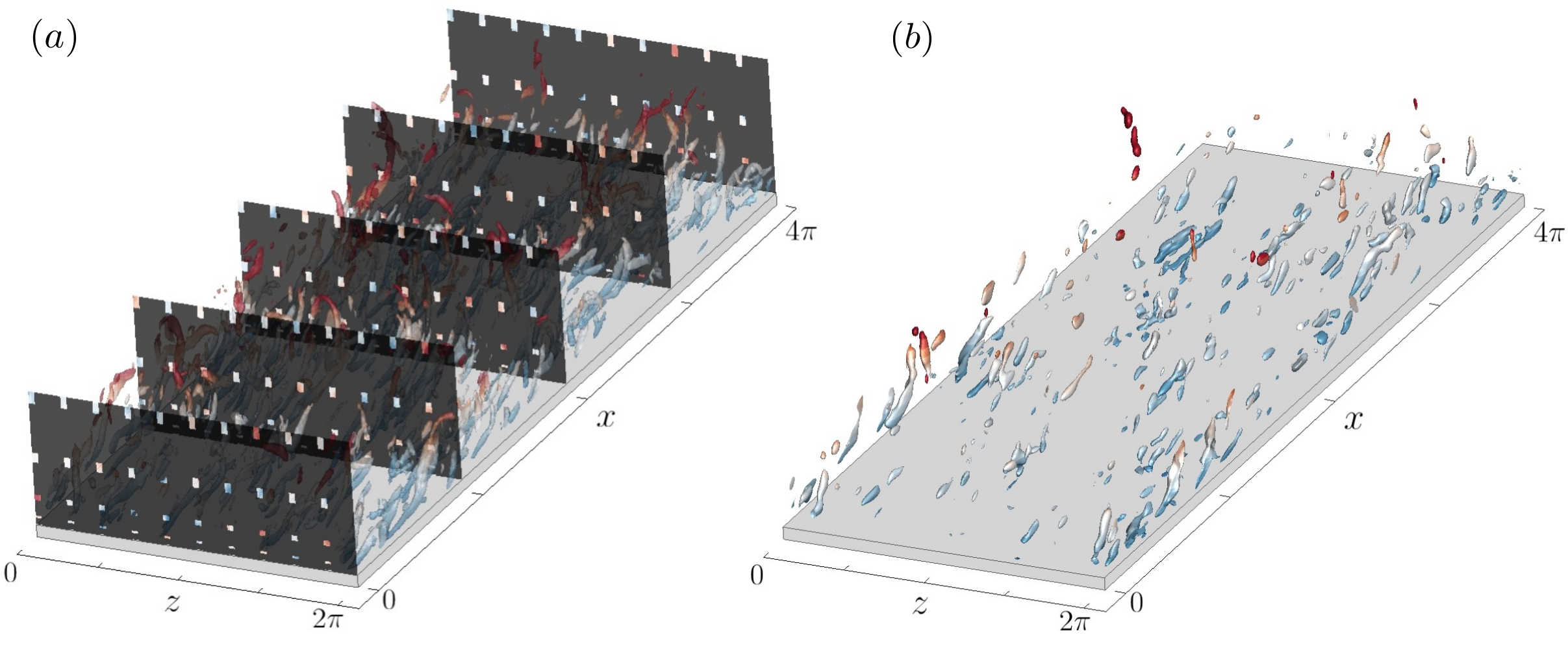}
    \caption{
    ($a$) True vortical structures visualized using the $\lambda_2$ vortex identification criterion,  with threshold $\lambda_2 = -2$. The isosurfaces are colored by wall-normal height, $y$.
    Pixels on the black cross-flow planes represent scarce observations, or sub-sampled velocity data. 
    ($b$) Vortical structures computed from interpolation of the observations.
    }
    \label{fig:data}
\end{figure}

In this section, we will attempt to reconstruct fully turbulent channel flow from sparse velocity data, and explore the critical data resolution that guarantees an accurate estimation.
In order to minimize uncertainties, have full control of the resolution of observations and be able to assess the accuracy of our reconstruction, we extracted our observations from an independent direct numerical simulation (DNS).  The flow in the channel was driven by a fixed pressure gradient at $Re_{\tau}=u_{\tau} h / \nu = 180$, where $u_{\tau}$ is the friction velocity and $h$ is half-channel height.
The observations are collected by sub-sampling the velocity field as shown in figure \ref{fig:data}$a$.
The observations were distributed in the streamwise, wall-normal and spanwise directions separated by $\Delta x_m^+ = 47$, $\Delta y^+_m \in [1.8, 24]$ and $\Delta z_m^+ = 28$, respectively, and their temporal resolution was $\delta t^+_m = 0.46$.  Taken all together, the observations are 1/4096 of the data resolution required for direct numerical simulations.
The estimation window was $T = 4.2$ ($T^+ = 50$), which was approximately one Lyapunov timescale of the flow \citep{nikitin2018characteristics}. 
This setup for observations is motivated by the trade-off between field of view and resolution in volume measurements in experiments, for example in particle image velocimetry (PIV).

Estimating the turbulent state by simply interpolating the available sparse observations is fruitless, especially if we are interested in the velocity gradients.  Figure \ref{fig:data}$b$ shows the outcome of such approach, where only a small portion of the vortical structures are captured in the interpolated flow field.  Naturally, the accuracy of the computed structures further deteriorates if the observations are contaminated by noise \citep{Abrahamson1995}.

\begin{figure}
    \centering
    \includegraphics[width = \textwidth]{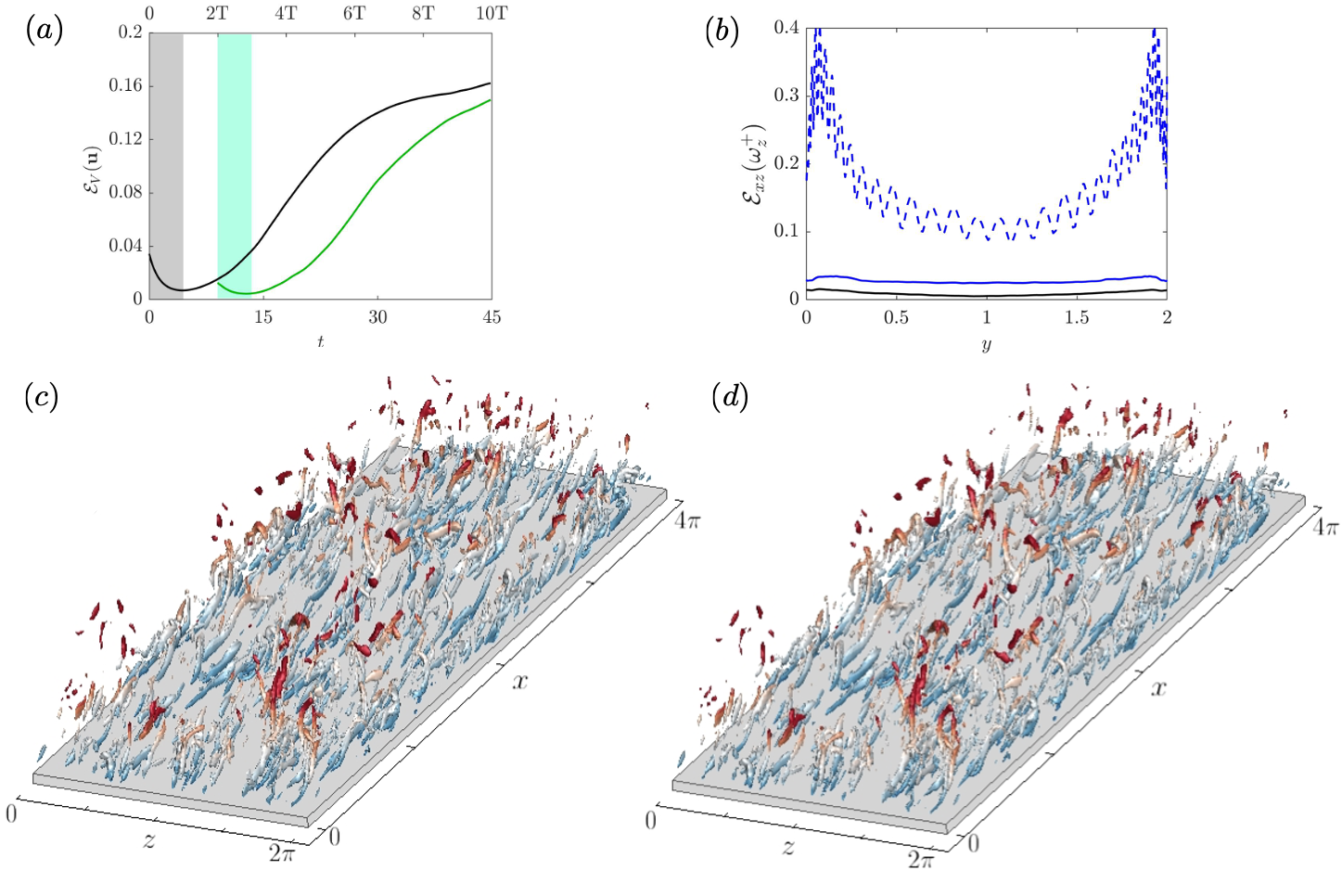}
    \caption{
    ($a$) Temporal evolution of volume-averaged error of the adjoint-variational estimated velocity field. Black line: initial condition at $t=0$ is estimated using data within $t \in [0,T]$; 
    Green line: flow state at $t=2T$ is estimated using data within $t \in [2T,3T]$.
    ($b$) Horizontally averaged error $\mathcal E_{xz}(\omega_z^+)$ in spanwise vorticity field at $t = T$. (Solid) Adjoint-variational estimation was performed using velocity data that is (black) noise-free and (blue) contaminated by 10\% noise relative to the local velocity.
    (Blue dashed) Vorticity is evaluated directly from interpolation of noisy velocity data. 
    Errors are normalized by mean vorticity $\langle \partial u / \partial y\rangle$ at the wall.
    ($c$,$d$) Estimated and true vortical structures at $t=T$, visualized using the $\lambda_2$ vortex identification criterion, with threshold $\lambda_2 = -2$ and colored by wall-normal distance $y$. 
    }
    \label{fig:field}
\end{figure}

Adjoint-variational data assimilation was performed in order to accurately estimate the flow state.
The first guess of the estimated initial condition was a simple interpolation of the observations, and we performed one hundred forward-adjoint loops.  The procedure reduced the cost function to $2.8\%$ of its initial value.
The accuracy of the predicted flow evolution can be evaluated by comparison to the hidden truth.
The volume-averaged estimation error is plotted in figure \ref{fig:field}$a$ (black line) as a function of time. 
Within $t \in [0,T]$ (gray region), the error decays monotonically and the flow more closely shadows the true trajectory in state space.  At the end of the assimilation window, $t=T$, the error becomes an order of magnitude smaller than at the initial time. 
This trend is interesting and can be explained with reference to the notion of adjoint chaos introduced above.  
Perturbations to the adjoint system amplify in backward time.  As such, when one considers the adjoint equations (\ref{eq:cont_adj_div}-\ref{eq:cont_adj_mom}), the forcing by the cost function at final time has the maximum potential for amplification during the backward evolution to the initial condition.  
More precisely, the disparity between model predictions and observations at late times have the most pronounced impact on the gradient direction for updating the initial condition.  
As a result, while the assimilated initial state is thus optimized to reproduce the time-history of observations, it places progressively higher emphasis on later ones.

As commonly known based on everyday experience with weather prediction, forecasts are progressively less accurate.  The same is true here: beyond the observation horizon, the errors in the estimated flow amplify although accuracy within $[T,2T]$ may be considered acceptable since it is commensurate with the levels during the assimilation horizon.  It is important to note that the reported levels of error are on the order of a percent.  While the levels of errors are much more restrictive than the public demands for weather forecasts, they are essential for accurate interpretation of flow experiments and probing the flow physics.  
The errors amplify exponentially due to the chaotic nature of the flow and finally saturate when the turbulence in the channel is entirely decorrelated from the true flow.  
Should new observations become available, adjoint-variational data assimilation can be performed again starting from the previous estimate of the flow.  In figure  \ref{fig:field}$a$, new observations were made available within $t \in [2T,3T]$ (light green region).  The data assimilation procedures were repeated, and was able to drive the estimated trajectory towards the true state, again (green line).   This figure nicely captures the influence of both forward and adjoint chaos:  The very small differences between the green and black curves at $t=T$ lead to significant differences at long times due to forward amplification of perturbations.  Meanwhile, the errors in the green curve decay during the assimilation horizon $t\in[2T, 3T]$ due to amplification of perturbations in backward time during the adjoint estimation.

The robustness of the algorithm for predicting vorticity from noisy observations is examined in figure \ref{fig:field}$b$.  For references, we are also plotting the errors from simple interpolation of the noisy observations (dashed curve) when the noise level is 10\%.  In this case, the errors in the near wall region reach approximately $30\%$ of the mean wall vorticity.  By contrast, the adjoint-variational estimation (black solid line) reduces the error to $4\%$ of the wall vorticity.
The reconstructed vortical structures are visualized in figure \ref{fig:field}$c$ and compared to the true flow in figure \ref{fig:field}$d$.
Although some of the small-scale structures are not perfectly captured, most of the estimated wall-attached and detached vortical structures are almost identical to the true flow.
This compelling estimation accuracy demonstrates the capacity of our algorithm to augment under-resolved turbulence data.

Intuitively, as the spatio-temporal spacing of observations is increased, reconstruction of the entire turbulent state becomes more difficult.
Through a systematic effort and numerous data assimilation experiments, we explored the critical resolution of observations that can enable an accurate reconstruction of the turbulence, with 90\% correlation coefficient between the assimilated and true states.  
Two important physical scales emerged as key factors: the Taylor microscale $\Lambda$ and the Lyapunov timescale $\tau_{\sigma}$.
The former is evaluated from the two-point velocity correlation, but perhaps more informative is its physical interpretation in the context of isotropic turbulence where it measures the distance traversed by a Kolmogorov eddy during its lifetime as it is swept by the root-mean-squared velocity.  This description can be reframed in backward time: An observation of a Kolmogorov eddy will be swept that distance during its adjoint lifetime. As a result, if observations are spaced at twice the Taylor microscale, their signal can be decoded to accurately reconstruct the entire volume. 
The Lyapunov timescale $\tau_{\sigma}$ is the e-folding time of an initial disturbance.
Even if the observation sites are closer than $2\Lambda$, the temporal sampling interval cannot exceed $\tau_{\sigma}$ or else Lyapunov amplification of errors will become dominant. 
In effect, the spatio-temporal data resolution must satisfy
\begin{equation}
   \label{eq:criteria_1}
   \Delta_m \lesssim 2 \Lambda, \quad \delta t_m \lesssim \tau_{\sigma},
\end{equation}
to guarantee an accurate turbulence reconstruction.
In the direction of mean advection, we can trade spatial and temporal resolutions of observations, 
\begin{equation}
    \label{eq:criteria_2}
    \delta t_m \lesssim 2 \Lambda / U_a, \quad \Delta x_m \lesssim U_a\tau_{\sigma}. 
\end{equation}
These criteria were examined in detail and supported by numerical experiments \citep{mengze2021} at $Re_\tau = 180$.  Future investigations at higher Reynolds numbers, with larger separation of scales, can further verify and refine these conditions.

The criteria discussed above hint to the notion of the domain of dependence of observations.  Intuition suggests that this domain depends on the type of observation, its placement in the flow and its separation in time from the initial condition.  
Absent any advection, we can picture that an isolated observation soon after the initial condition depends on a small spherical region due to diffusion of nearby effects, and that region is larger for later observations. 
When mean advection is present, we can picture an plume-like region upstream of the observation site, which marks all the potential locations where early events could have reached the sensor.  
This intuitive picture will be more precise using the adjoint field in the next section.

\subsection{Domain of dependence and Hessian matrix}
\label{sec:hessian}

The domain of dependence of a sensor can be mathematically defined by its sensitivity to the initial flow state.  This idea is best explored in terms of variations, specifically whether and to what extent a small change in the initial flow state affects the observation at the sensing location and time.  We can therefore consider a small perturbation to the true flow $\boldsymbol u^{\prime} = \boldsymbol u - \boldsymbol u^R$, which is governed by the 
linearized Navier-Stokes equations $\boldsymbol u^{\prime} = \mathcal A \boldsymbol u_0^{\prime}$, and evaluate its associated influence on the observation, $\mathcal M(\boldsymbol u) - \mathcal M(\boldsymbol u^R) = \mathcal M(\boldsymbol u^{\prime})$ for a linear observation operator.   
The same deviation in the observation can also be evaluated using the Lagrangian duality relation, 
\begin{eqnarray}
    \label{eqn:duality}
    \mathcal M(\boldsymbol u^{\prime}) &=& \left[\boldsymbol{u}^{\prime}(t_m), \boldsymbol{\phi}\left(\boldsymbol{x}_m\right)\right] 
    = \left[ \mathcal{A}\boldsymbol{u}^{\prime}_0, \boldsymbol{\phi} \right] \nonumber \\
    &=& \left[ \boldsymbol{u}^{\prime}_0, \mathcal{A}^{\dag}\boldsymbol{\phi} \right] = \left[\boldsymbol{u}^{\prime}_0, \boldsymbol{u}^{\dag}(\tau = t_m;\boldsymbol{x}_m,t_m)\right],
\end{eqnarray}
where $\boldsymbol \phi$ is the observation kernel, and $\mathcal A^{\dag}$ is the adjoint linearized Navier-Stokes operator.
Here we focus on an individual observation from a sensor placed at $\boldsymbol x_m$ recorded at the time instant $t_m$, which is the elementary building block of the cost function.
For this instantaneous observation, the inner product in equation \ref{eqn:duality} only involves spatial integration.
As the final expression in equation \ref{eqn:duality} demonstrates, the sensitivity of observation to the initial state is quantified by the adjoint field $\boldsymbol {u}^{\dag}(\tau = t_m;\boldsymbol{x}_m,t_m)$.
The support of $\boldsymbol {u}^{\dag}$ represents the \emph{domain of dependence} of an observation recorded at $\left(\boldsymbol {x}_m, t_m\right)$.

\begin{figure}
    \centering
    \includegraphics[width = 0.9\textwidth]{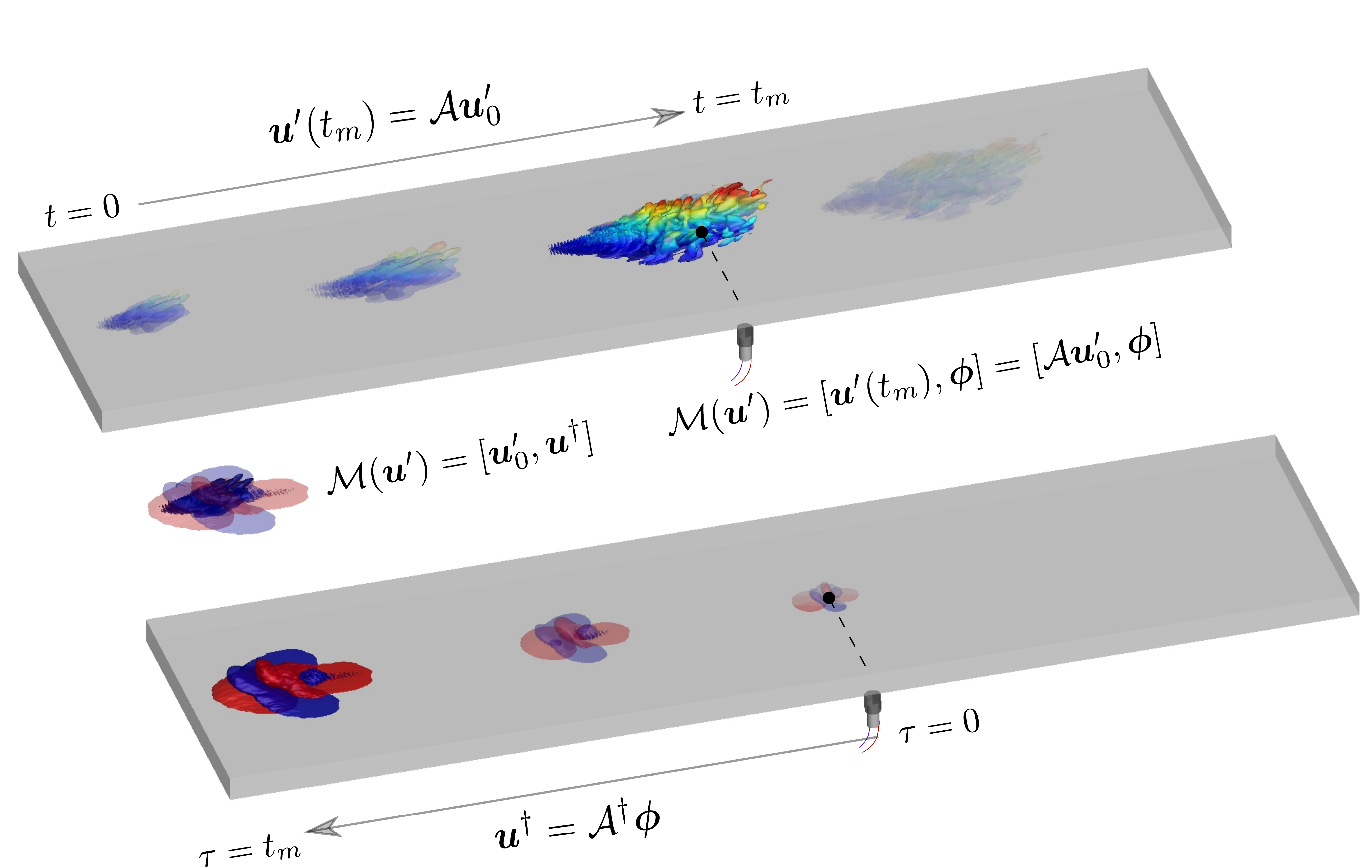}
    \caption{
    Schematic of forward-adjoint duality (\ref{eqn:duality}) in turbulent channel flow.
    Top: Forward evolution of an initial perturbation, with isosurfaces showing the streamwise velocity $u^{\prime}$, colored by wall-normal distance $y$.
    The associated impact on the observation is evaluated at the wall at $t^+ = t_m^+ = 20$.
    Bottom: Backward evolution of the adjoint field starting from an impulse at the wall, up to $\tau = t_m$, where isosurfaces mark (red) positive and (blue) negative adjoint streamwise velocities $u^{\dag}$.
    Modification to the observation due to any initial forward perturbation can be efficiently computed from its inner product with this single adjoint field at $t = 0$ ($\tau = t_m$).
    }
    \label{fig:DoD}
\end{figure}

The Lagrangian duality (\ref{eqn:duality}) is illustrated in figure \ref{fig:DoD}.
In this example, the observation kernel $\boldsymbol \phi$ corresponds to evaluating the streamwise wall stress at $(\boldsymbol x_m,t_m)$ from the velocity field.
The top panel shows the forward approach:
we perturb every velocity component at every point in the initial condition separately, and evaluate the corresponding deviation in the wall observation.
This approach is therefore computationally expensive because it requires $(nx \times ny \times nz \times 3)$ forward simulations, each evolved until $t=t_m$.
The last equality in (\ref{eqn:duality}) is significantly more efficient. The adjoint field $\boldsymbol{u}^{\dag}(\tau = t_m;\boldsymbol{x}_m,t_m)$ is obtained from a single simulation of the adjoint equations backward in time,
\begin{eqnarray}
	\label{eq:impulse_adj_div}
    \nabla \cdot \boldsymbol u^{\dag} &=& 0 \\
	\label{eq:impulse_adj_mom}
    \frac{\partial \boldsymbol u^{\dag}}{\partial \tau} - \boldsymbol u^R \cdot \nabla \boldsymbol u^{\dag} + (\nabla \boldsymbol u^R) \cdot \boldsymbol u^{\dag} &=& \nabla p^{\dag} + \frac{1}{Re} \nabla^2 \boldsymbol u^{\dag}, \\
    \label{eq:impulse_adj_IC}
    \boldsymbol u^{\dag}(\tau = 0) &=& \boldsymbol \phi(\boldsymbol x_m).
\end{eqnarray}
Note that the adjoint equations (\ref{eq:impulse_adj_div}-\ref{eq:impulse_adj_IC}) are different from the earlier adjoint system (\ref{eq:cont_adj_div}-\ref{eq:cont_adj_mom}) that was driven by a time-dependent forcing from the cost.  Here the adjoint field is initiated at $\tau = 0$ from an impulse at the sensor position (Eq.~\ref{eq:impulse_adj_IC}), and its evolution uses the true \emph{reference} state $\boldsymbol u^R$ (Eq.~\ref{eq:impulse_adj_mom}).  The resulting adjoint field thus quantifies the sensitivity of wall observations to deviation from the \emph{reference} initial state.  Any such deviation $\boldsymbol u_0^{\prime}$ affects the observations at $t_m$ if and only if $\boldsymbol u_0^{\prime}$ is non-zero within the support of $\boldsymbol u^{\dag}$, as shown in the bottom panel of figure \ref{fig:DoD}.

We will focus the application of the above formulation on wall observations in turbulent channel flow due to their theoretical significance.  The generation of all the interior vorticity in the channel can be traced back to the wall \citep{Eyink2020_theory,Eyink2020_channel}.
The converse problem, specifically whether the entire initial turbulent state can be decoded from wall signals, has not been addressed comprehensively. 
Previous efforts using a variety of approaches \citep{bewley2004skin,hasegawa2016estimation,encinar2019logarithmic,mengze2021} have demonstrated that the near-wall turbulence and only the outer large-scale structures can be reconstructed from the wall stresses.
By quantifying the domain of dependence of wall sensors, we attempt to demystify the difficulty of flow reconstruction from wall observations.

The domain of dependence of a wall stress $\nu \partial u/\partial y$ sensor is in fact the adjoint field in figure \ref{fig:DoD} (bottom panel).   Unlike the forward evolution of an impulse, the adjoint structures are advected upstream of the observation point and expand as a function of reverse time.  This behavior was introduced conceptually in the previous section, and is here evaluated using the adjoint equations. 
The isosurfaces of $u^{\dag}$ are symmetric in the spanwise direction because disturbance to the forward $u$ velocity on either side of the measurement cannot be distinguished.
At $\tau = t_m$ $(t=0)$, the adjoint field $\boldsymbol u^{\dag}(\tau = t_m; \boldsymbol x_m,t_m)$ measures the sensitivity of wall observation to initial velocity at every point.
The upstream orientation and pancake-like shape of the adjoint isosurfaces in figure \ref{fig:DoD} implies that the wall observation at $t_m^+ = 20$ is most sensitive to upstream and near-wall perturbations at the initial time.

Since channel-flow turbulence is statistically homogeneous in the horizontal dimensions, observations can be sampled over the entire wall. The cost function associated with small deviation in the observation becomes,
\begin{equation}
    \label{eq:cost_wall}
    \mathcal J(\boldsymbol u_0^{\prime}; t_m) 
    = \frac 1{2S} \int_S \left(\frac{\partial u^{\prime}}{\partial y}\right)^2 dx_m dz_m
    =\frac 1{2S} \int_S [\boldsymbol u_0^{\prime},\boldsymbol u^{\dag}]^2 dx_m dz_m
\end{equation}
where the last equality invoked the Lagrangian duality (\ref{eqn:duality}) and $S$ is the area of the wall.
The cost function (\ref{eq:cost_wall}) is quadratic in $\boldsymbol u_0^{\prime}$, and its gradient vanishes at the true solution ($\boldsymbol u_0^{\prime} = \boldsymbol 0$).
As a result, the difficulty of flow reconstruction near the true solution can be characterized by the Hessian matrix, 
\begin{equation}
    \label{eq:hessian_derive}
    \mathcal H(\boldsymbol x_1, \boldsymbol x_2; t_m) \equiv \frac{\partial \mathcal J}{\partial \boldsymbol u_0^{\prime} \partial \boldsymbol u_0^{\prime}} = \frac {1}{S} \int_S \boldsymbol u^{\dag} \boldsymbol u^{\dag} dx_m dz_m.
\end{equation}
Written explicitly, the Hessian has the form,
\begin{equation}
    \label{eq:hessian_physical}
    \mathcal H_{ij}(\boldsymbol x_1, \boldsymbol x_2; t_m) = \frac {1}{S} \int_S u^{\dag}_i (\boldsymbol x_1, \tau = t_m; \boldsymbol x_m,t_m) u^{\dag}_j (\boldsymbol x_2, \tau = t_m; \boldsymbol x_m,t_m) dx_m dz_m,
\end{equation}
which demonstrates that the Hessian matrix is the auto-correlation of the adjoint field, which corresponds to the observability Gramian in control theory \citep{Rowley2005gramian}.
Here, the Hessian analysis characterizes the difficulty of interpreting observations in order to reconstruct the flow state near optimality, and our particular interest is in non-linear time-dependent turbulent flows.

Eigen-analysis of the Hessian matrix can be viewed from two perspectives:
(i) The leading eigenfunctions describe the directions that adjoint-variational method will most effectively target during the data assimilation procedure. In other words, these flow structures will be the easiest to reconstruct from observations.
The associated eigenvalues quantify the curvature of the cost function along the corresponding eigen-direction. 
(ii) Eigen-analysis of the Hessian matrix is equivalent to a proper-orthogonal decomposition (POD) of the adjoint field $\boldsymbol u^{\dag}$.
As such, the leading eigenfunctions are the dominant shapes of the domain of dependence.
In the forward sense, these eigenfunctions are also the optimal perturbations that will lead to the most significant change in wall signals at $t_m$, and the corresponding sensitivity is quantified by the eigenvalues.

\begin{figure}
    \centering
    \includegraphics[width = 0.9\textwidth]{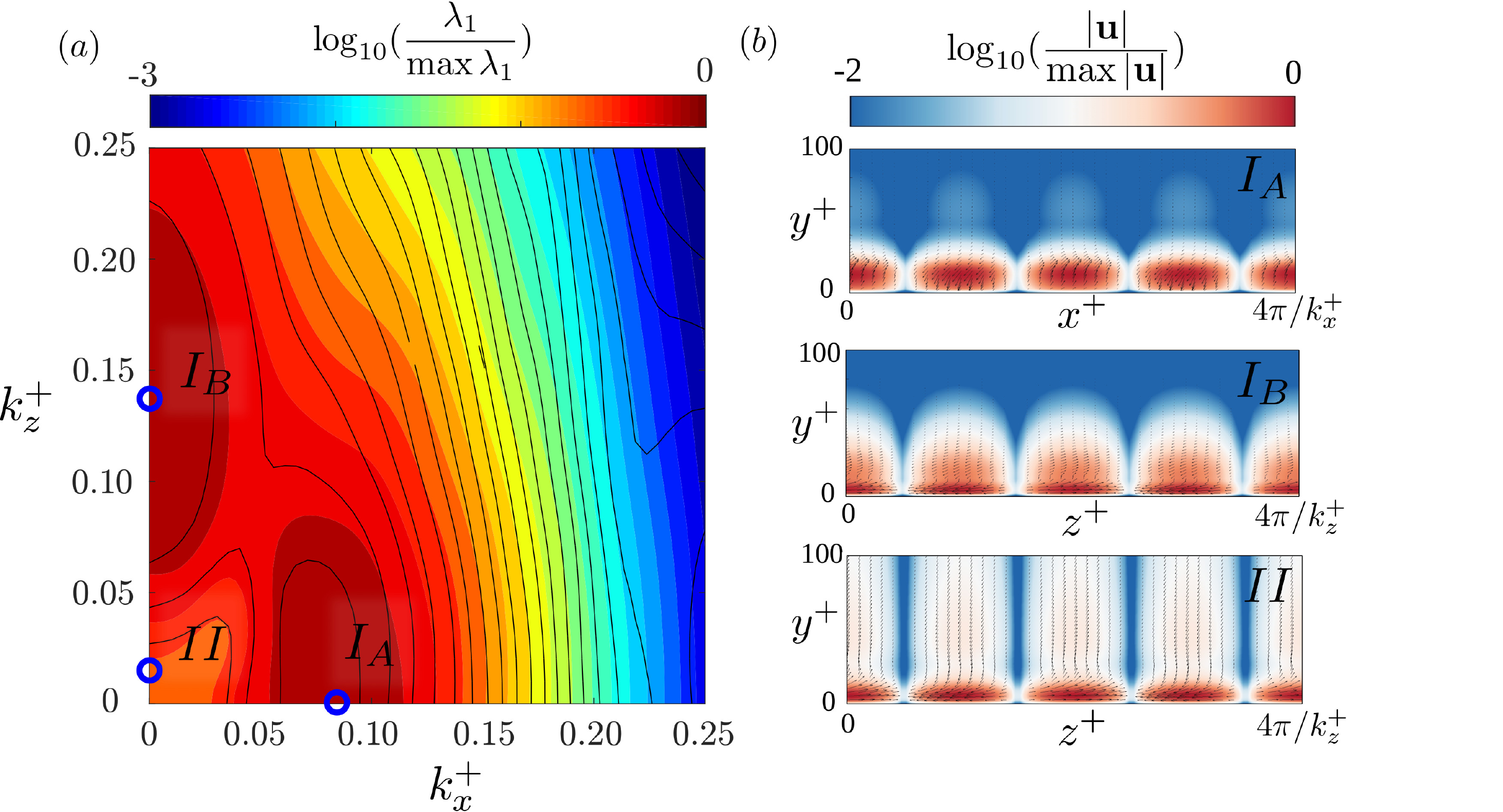}
    \caption{
    ($a$) The largest eigenvalue of the Hessian matrix $\hat {\mathcal H}(k_x,k_z;t_m)$ for observing $\partial u / \partial y|_{\mathrm{wall}}$ instantaneously at $t_m^+ = 20$. 
    Color and line contours are for $Re_{\tau} = 180$ and $590$, respectively.
    The eigenvalues are normalized by the supremum. 
    ($b$) Visualizations of the leading eigenfunctions of the Hessian at selected wavenumber pairs:
    $(k_x^+, k_z^+)$ = (0.08, 0), (0, 0.14), (0, 0.02) for modes $I_A$, $I_B$ and $II$.
    Contours show the velocity magnitude and arrows are in-plane velocity components.
    }
    \label{fig:hessian}
\end{figure}

A sample eigenspectrum and eigenfunctions of the Hessian for observing $\partial u/\partial y|_{\mathrm{wall}}$ is shown in figure \ref{fig:hessian}.
Exploiting periodicity in the horizontal $x$-$z$ plane, the Hessian is expressed in Fourier space,
\begin{equation}
    \label{eq:hessian_fourier}
    \hat {\mathcal H}(k_x,k_z;t_m) = \frac {1}{S} \int_S \hat {\boldsymbol u}^{\dag} (k_x,k_z, \tau = t_m; \boldsymbol x_m,t_m) \hat {\boldsymbol u}^{\dag} (k_x,k_z, \tau = t_m; \boldsymbol x_m,t_m) dx_m dz_m.
\end{equation}
The largest eigenvalue of $\hat{\mathcal H}(k_x,k_z;t_m)$ at each wavenumber pair is plotted in figure \ref{fig:hessian}$a$.
Two Reynolds numbers $Re_{\tau} = 180$ and $590$ are considered, and their eigenspectra (color and line contours) coincide when the wavenumber are scaled in viscous units. 
The maximum eigenvalue decays with increasing magnitude of the horizontal wavenumber vector $(k_x^+, k_z^+)$, and the separation of eigenvalues over three orders of magnitudes demonstrates the ill-conditioned nature of the Hessian. 
Two local peaks marked $I_A$ and $I_B$ correspond to spanwise and streamwise rolls.
Another important point is marked $II$, which represents the energetic large-scale structures with $\lambda_z^+ = \mathcal O(300)$.
The mode shapes of $I_A$, $I_B$ and $II$ are plotted in figure \ref{fig:hessian}$b$.

Within the considered time horizon, $t_m^+ = 20$, the initial disturbances that can most effectively influence wall measurements have large wavenumbers, and the corresponding eigenvectors are clustered near the wall (modes ${I_A}$ and $I_B$ in figure \ref{fig:hessian}$b$).
These modes are not energetic in channel-flow turbulence, but are nonetheless the directions that adjoint-variational data assimilation will attempt to reconstruct first.  
Another important remark is that, beyond the buffer layer, most eigenfunctions are vanishing \citep[see][for more details]{Qi2021}, which implies that wall observations are not very sensitive to the turbulence far from the wall.  A very important exception is in a small region near mode $II$, where the eigenfunctions remain finite in the core of the channel. These modes represent the sensitivity of wall signals to outer large-scale structures. 
This sensitivity arises because these streamwise-elongated motions are immune to shear sheltering \citep{hunt1999perturbed,zaki2009shear}, have a footprint in the near-wall region and modulate the near-wall structures \citep{mathis2009large,abe2004very,hwang2016inner,you2019tbl}.  
As a result, they have a signature encoded in the wall stress.  
The analysis of the Hessian eigenspectrum thus provides a rigorous explanation of the empirical findings by earlier studies \citep{hasegawa2016estimation,encinar2019logarithmic,mengze2021}: while wall observations can be decoded to discover all the turbulent scales near the wall, only the outer large-scale motions can be accurately estimated from wall stresses.

The present discussion has focused on an instantaneous observation, at $t_m^+ = 20$.
In practice, observations are accumulated over a time horizon that may extend over a longer duration.
Due to the Lyapunov behaviour of the \emph{adjoint} system, the adjoint field amplifies exponentially in reverse time at the same rate as the forward system. 
The notion of the adjoint chaos is, however, different from the familiar forward problem: 
To start, the initial guess of the flow state in a data assimilation procedure is generally far from the truth, and the associated gradient of the cost function is determined by the adjoint system.
The exponential amplification of the adjoint in backward time can lead to a very large gradient, which imposes a severe restriction on the step size in gradient descent optimization.
Even if the initial estimate is near optimality, the Hessian matrix, which is represented by the adjoint field starting from an observation kernel, has exponentially amplifying eigenvalues with observation time. 
This property translates into a cost function landscape with extreme curvatures, which can thwart the search for the global optimum.
More subtle but very important is an increase in the separation between the largest and smallest eigenvalues of the Hessian, and hence an increase of its condition number. As a result, solving the inverse problem becomes progressively more ill conditioned and errors in observations can strongly obscure the reconstruction of the initial state.
Effective strategies to address these difficulties include limiting the assimilation horizon based on knowledge of the Lyapunov timescale \citep{li2020,mengze2021}, and for longer horizons adopting a sliding window \citep{Chandramouli2020} or a cycling scheme \citep{Fisher2012}. These techniques should also be combined with strategies for optimizing sensor placement and weighting \citep{mons2019kriging,buchta2021envar} in order to maximize the accuracy of predictions.

\section{Conclusion}
\label{sec:conclusion}

Data assimilation combines experimental observations with numerical simulations in order to achieve high-fidelity predictions of the entire flow state.
The data-infused simulations improve the accuracy of the computations and reduce the uncertainty of predictions.
At the same time, experimental measurements are augmented with non-intrusive access to a much higher resolution than the sensor data.  
These benefits have been demonstrated by applying adjoint-variational data assimilation in order to reconstruct a nonlinearly saturated flow and wall turbulence, from limited observations.  
When the flow configuration admits multiple states, such as in circular Couette flow, the observed one may originate from a complicated history and particulars of the experimental setup. Data assimilation can successfully and accurately reconstruct the true state from sparse observations, such as wall stress measurements, without the requirement of matching the detailed experimental setup and protocol \citep{mengze2019discrete}. 
In wall turbulence, starting from sparse velocity data, adjoint-variational assimilation provided reconstructions of the full velocity and vorticity fields that are nearly perfectly correlated with the truth and robust to observation noise \citep{mengze2021}. 

The importance of data assimilation extends beyond augmenting observations.  
In particular, adjoint-variational methods provide a unique perspective on flow physics that arises from the process of decoding observations.
The adjoint field quantifies the sensitivity of the observation kernel to changes in the flow state, and hence the domain of dependence of the sensor.  
This view motivates a shift in focus from clustering measurements in order to locally observe all the scales to optimally distributing observations in order to ensure sensitivity to the entire state. 
In order to guarantee accurate reconstruction of the flow, a critical data resolution must be satisfied, whereby the domains of dependence of adjacent observation sites overlap.  This resolution was shown to be on the order of the Taylor microscale of turbulence.

The search for the true flow is performed by minimizing a cost function that is defined in terms of the disparity between model predictions and observations. Success must therefore contend with the landscape of this cost function, which is mountainous for nonlinear chaotic systems.  We have shown an example of the cost landscape explicitly using the Lorenz system\textemdash a practically impossible task in high dimensions such as for turbulence.  
Luckily the adjoint system provides a solution \citep{Qi2021}: At optimality, the Hessian matrix of the cost can be computed relatively efficiently from the adjoint field, and its eigen-decomposition provides important information. 
The leading eigenfunctions are the flow structures that have the most impact on observations. 
The eigenvalues are the curvatures of the cost landscape along the associated eigen-directions, and describe the difficulty of approaching optimality along those directions.
As such, from the perspective of data assimilation, or the inverse problem, the leading eigenfunction are the most accurate to decode from observations.

The above ideas were applied to the interpretation of observations of the wall stress in turbulent channel flow.
Within a short time horizon, these observations are most sensitive to near-wall small-scale perturbations, and strongly affected by outer large-scale structures.
At longer observation time horizons, the sensitivity amplifies exponentially due to the chaotic nature of the adjoint system, leading to a more mountainous cost function, larger gradient, and a deteriorating ill-conditioned Hessian.
Therefore, the inverse problem becomes progressively more difficult to solve, and errors in observations can strongly contaminate the estimated initial state.
These difficulties restrict assimilation time horizon to the order of the Lyapunov timescale.

The difficulty of turbulence reconstruction can be alleviated by designing optimal measurement layouts, which can be guided by the Hessian analysis.
The weighting of observations and spatio-temporal placement of sensors can be optimized to mitigate undesirable features of the cost landscape. 
For example, in transitional or turbulent flows earlier observations can be favorably weighted in order to reduce extreme curvature, and thus improve the estimation accuracy of the initial state.  In turbulent flows, the sensor placement can be designed to mitigate the poor condition of the Hessian \citep{buchta2021envar}.  
These strategies have been applied to prediction of high-speed boundary layer transition and scalar source reconstruction in turbulent environments \citep{wang_hasegawa_zaki_2019,mons2019kriging}.

At much higher Reynolds numbers, the challenges examined herein become more severe and new ones arise.
The higher Lyapunov exponent will further restrict the duration of the assimilation window.
In addition, a larger number of observations, or sensors, will be required at higher Reynolds numbers, which can render estimation of the entire flow field prohibitive both experimentally and computationally. Measurements can therefore be concentrated in a small region of interest, and the data assimilation can reconstruct the flow within this sub-domain along with the effective boundary conditions due to the influence of the larger configuration. Finally, turbulence modeling may be required in the forward and adjoint equations, which introduces uncertainty with respect to model parameters. All these considerations are exciting future avenues of inquiry. 

Furthermore, the results from the adjoint-variational approach provide a benchmark for assessing emerging data-assimilation techniques, for example recent machine learning strategies that integrate data and physics-based, or equations, constraints.  
Notable among the growing suite of these methods are physics-informed neural networks \citep{Raissi2019PINN}, evolutionary deep neural networks \citep[EDNN which is pronounced Eden, see][]{Du2021EDNN}, and Deep Operator Networks \citep[DeepONet, see ][]{PCDL2021DeepONet}. 
A systematic comparison of adjoint-variational and machine-learning data assimilation techniques is currently under investigation, and future efforts may combine the two classes of methods in multi-fidelity data-assimilation strategies.


\par\bigskip
\noindent
\textbf{Acknowledgements.} 
The authors are grateful to Dr.\,David Buchta and Dr.\,Qi Wang for their contributions and perspectives on interpretation of adjoint fields and the Hessian analysis of chaotic systems.
The authors acknowledge financial support from the Office of Naval Research (N00014-20-1-2715,  N00014-21-1-2375) and the Air Force Office of Scientific Research (FA9550-19-1-0230). Computational resources were provided by the Maryland Advanced Research Computing Center (MARCC).

\noindent
\textbf{Declaration of interests.} 
The authors report no conflict of interest.
 

\bibliography{Manuscript}

\begin{thebibliography}{53}%
\makeatletter
\providecommand \@ifxundefined [1]{%
 \@ifx{#1\undefined}
}%
\providecommand \@ifnum [1]{%
 \ifnum #1\expandafter \@firstoftwo
 \else \expandafter \@secondoftwo
 \fi
}%
\providecommand \@ifx [1]{%
 \ifx #1\expandafter \@firstoftwo
 \else \expandafter \@secondoftwo
 \fi
}%
\providecommand \natexlab [1]{#1}%
\providecommand \enquote  [1]{``#1''}%
\providecommand \bibnamefont  [1]{#1}%
\providecommand \bibfnamefont [1]{#1}%
\providecommand \citenamefont [1]{#1}%
\providecommand \href@noop [0]{\@secondoftwo}%
\providecommand \href [0]{\begingroup \@sanitize@url \@href}%
\providecommand \@href[1]{\@@startlink{#1}\@@href}%
\providecommand \@@href[1]{\endgroup#1\@@endlink}%
\providecommand \@sanitize@url [0]{\catcode `\\12\catcode `\$12\catcode
  `\&12\catcode `\#12\catcode `\^12\catcode `\_12\catcode `\%12\relax}%
\providecommand \@@startlink[1]{}%
\providecommand \@@endlink[0]{}%
\providecommand \url  [0]{\begingroup\@sanitize@url \@url }%
\providecommand \@url [1]{\endgroup\@href {#1}{\urlprefix }}%
\providecommand \urlprefix  [0]{URL }%
\providecommand \Eprint [0]{\href }%
\providecommand \doibase [0]{https://doi.org/}%
\providecommand \selectlanguage [0]{\@gobble}%
\providecommand \bibinfo  [0]{\@secondoftwo}%
\providecommand \bibfield  [0]{\@secondoftwo}%
\providecommand \translation [1]{[#1]}%
\providecommand \BibitemOpen [0]{}%
\providecommand \bibitemStop [0]{}%
\providecommand \bibitemNoStop [0]{.\EOS\space}%
\providecommand \EOS [0]{\spacefactor3000\relax}%
\providecommand \BibitemShut  [1]{\csname bibitem#1\endcsname}%
\let\auto@bib@innerbib\@empty
\bibitem [{\citenamefont {Yeung}\ \emph {et~al.}(2018)\citenamefont {Yeung},
  \citenamefont {Sreenivasan},\ and\ \citenamefont {Pope}}]{Yeung2018dns}%
  \BibitemOpen
  \bibfield  {author} {\bibinfo {author} {\bibfnamefont {P.~K.}\ \bibnamefont
  {Yeung}}, \bibinfo {author} {\bibfnamefont {K.~R.}\ \bibnamefont
  {Sreenivasan}},\ and\ \bibinfo {author} {\bibfnamefont {S.~B.}\ \bibnamefont
  {Pope}},\ }\bibfield  {title} {\bibinfo {title} {Effects of finite spatial
  and temporal resolution in direct numerical simulations of incompressible
  isotropic turbulence},\ }\href@noop {} {\bibfield  {journal} {\bibinfo
  {journal} {Phys. Rev. Fluids}\ }\textbf {\bibinfo {volume} {3}},\ \bibinfo
  {pages} {064603} (\bibinfo {year} {2018})}\BibitemShut {NoStop}%
\bibitem [{\citenamefont {Lee}\ and\ \citenamefont {Moser}(2015)}]{Lee2015dns}%
  \BibitemOpen
  \bibfield  {author} {\bibinfo {author} {\bibfnamefont {M.}~\bibnamefont
  {Lee}}\ and\ \bibinfo {author} {\bibfnamefont {R.~D.}\ \bibnamefont
  {Moser}},\ }\bibfield  {title} {\bibinfo {title} {Direct numerical simulation
  of turbulent channel flow up to $\mathit{Re}_{{\it\tau}}\approx 5200$},\
  }\href@noop {} {\bibfield  {journal} {\bibinfo  {journal} {J.~Fluid Mech.}\
  }\textbf {\bibinfo {volume} {774}},\ \bibinfo {pages} {395} (\bibinfo {year}
  {2015})}\BibitemShut {NoStop}%
\bibitem [{\citenamefont {Marxen}\ \emph {et~al.}(2014)\citenamefont {Marxen},
  \citenamefont {Iaccarino},\ and\ \citenamefont {Magin}}]{marxen2014direct}%
  \BibitemOpen
  \bibfield  {author} {\bibinfo {author} {\bibfnamefont {O.}~\bibnamefont
  {Marxen}}, \bibinfo {author} {\bibfnamefont {G.}~\bibnamefont {Iaccarino}},\
  and\ \bibinfo {author} {\bibfnamefont {T.~E.}\ \bibnamefont {Magin}},\
  }\bibfield  {title} {\bibinfo {title} {Direct numerical simulations of
  hypersonic boundary-layer transition with finite-rate chemistry},\
  }\href@noop {} {\bibfield  {journal} {\bibinfo  {journal} {J.~Fluid Mech.}\
  }\textbf {\bibinfo {volume} {755}},\ \bibinfo {pages} {35} (\bibinfo {year}
  {2014})}\BibitemShut {NoStop}%
\bibitem [{\citenamefont {Esteghamatian}\ and\ \citenamefont
  {Zaki}(2020)}]{esteghamatian_2021}%
  \BibitemOpen
  \bibfield  {author} {\bibinfo {author} {\bibfnamefont {A.}~\bibnamefont
  {Esteghamatian}}\ and\ \bibinfo {author} {\bibfnamefont {T.~A.}\ \bibnamefont
  {Zaki}},\ }\bibfield  {title} {\bibinfo {title} {Viscoelasticity and the
  dynamics of concentrated particle suspension in channel flow},\ }\href
  {https://doi.org/10.1017/jfm.2020.525} {\bibfield  {journal} {\bibinfo
  {journal} {J.~Fluid Mech.}\ }\textbf {\bibinfo {volume} {901}},\ \bibinfo
  {pages} {A25} (\bibinfo {year} {2020})}\BibitemShut {NoStop}%
\bibitem [{\citenamefont {Lee}\ and\ \citenamefont {Zaki}(2017)}]{lee_2017}%
  \BibitemOpen
  \bibfield  {author} {\bibinfo {author} {\bibfnamefont {S.~J.}\ \bibnamefont
  {Lee}}\ and\ \bibinfo {author} {\bibfnamefont {T.~A.}\ \bibnamefont {Zaki}},\
  }\bibfield  {title} {\bibinfo {title} {Simulations of natural transition in
  viscoelastic channel flow},\ }\href {https://doi.org/10.1017/jfm.2017.198}
  {\bibfield  {journal} {\bibinfo  {journal} {J.~Fluid Mech.}\ }\textbf
  {\bibinfo {volume} {820}},\ \bibinfo {pages} {232–262} (\bibinfo {year}
  {2017})}\BibitemShut {NoStop}%
\bibitem [{\citenamefont {Seo}\ \emph {et~al.}(2013)\citenamefont {Seo},
  \citenamefont {Vedula}, \citenamefont {Abraham},\ and\ \citenamefont
  {Mittal}}]{Seo2013cardio}%
  \BibitemOpen
  \bibfield  {author} {\bibinfo {author} {\bibfnamefont {J.~H.}\ \bibnamefont
  {Seo}}, \bibinfo {author} {\bibfnamefont {V.}~\bibnamefont {Vedula}},
  \bibinfo {author} {\bibfnamefont {T.}~\bibnamefont {Abraham}},\ and\ \bibinfo
  {author} {\bibfnamefont {R.}~\bibnamefont {Mittal}},\ }\bibfield  {title}
  {\bibinfo {title} {Multiphysics computational models for cardiac flow and
  virtual cardiography},\ }\href@noop {} {\bibfield  {journal} {\bibinfo
  {journal} {Int. J. Num. Meth. Biomed. Eng.}\ }\textbf {\bibinfo {volume}
  {29}},\ \bibinfo {pages} {850} (\bibinfo {year} {2013})}\BibitemShut
  {NoStop}%
\bibitem [{\citenamefont {Nicolaou}\ and\ \citenamefont
  {Zaki}(2013)}]{nicolaou2013}%
  \BibitemOpen
  \bibfield  {author} {\bibinfo {author} {\bibfnamefont {L.}~\bibnamefont
  {Nicolaou}}\ and\ \bibinfo {author} {\bibfnamefont {T.}~\bibnamefont
  {Zaki}},\ }\bibfield  {title} {\bibinfo {title} {Direct numerical simulations
  of flow in realistic mouth–throat geometries},\ }\href
  {https://doi.org/https://doi.org/10.1016/j.jaerosci.2012.10.003} {\bibfield
  {journal} {\bibinfo  {journal} {J. Aerosol Sci.}\ }\textbf {\bibinfo {volume}
  {57}},\ \bibinfo {pages} {71} (\bibinfo {year} {2013})}\BibitemShut {NoStop}%
\bibitem [{\citenamefont {Yang}\ and\ \citenamefont
  {Meneveau}(2016)}]{Yang2016inflow}%
  \BibitemOpen
  \bibfield  {author} {\bibinfo {author} {\bibfnamefont {X.~I.}\ \bibnamefont
  {Yang}}\ and\ \bibinfo {author} {\bibfnamefont {C.}~\bibnamefont
  {Meneveau}},\ }\bibfield  {title} {\bibinfo {title} {Recycling inflow method
  for simulations of spatially evolving turbulent boundary layers over rough
  surfaces},\ }\href@noop {} {\bibfield  {journal} {\bibinfo  {journal}
  {J.~Turbul.}\ }\textbf {\bibinfo {volume} {17}},\ \bibinfo {pages} {75}
  (\bibinfo {year} {2016})}\BibitemShut {NoStop}%
\bibitem [{\citenamefont {Wu}(2017)}]{Wu2017inflow}%
  \BibitemOpen
  \bibfield  {author} {\bibinfo {author} {\bibfnamefont {X.}~\bibnamefont
  {Wu}},\ }\bibfield  {title} {\bibinfo {title} {Inflow turbulence generation
  methods},\ }\href@noop {} {\bibfield  {journal} {\bibinfo  {journal} {Annu.
  Rev. Fluid Mech.}\ }\textbf {\bibinfo {volume} {49}},\ \bibinfo {pages} {23}
  (\bibinfo {year} {2017})}\BibitemShut {NoStop}%
\bibitem [{\citenamefont {Huisman}\ \emph {et~al.}(2014)\citenamefont
  {Huisman}, \citenamefont {Van Der~Veen}, \citenamefont {Sun},\ and\
  \citenamefont {Lohse}}]{Lohse2014TC}%
  \BibitemOpen
  \bibfield  {author} {\bibinfo {author} {\bibfnamefont {S.~G.}\ \bibnamefont
  {Huisman}}, \bibinfo {author} {\bibfnamefont {R.~C.}\ \bibnamefont {Van
  Der~Veen}}, \bibinfo {author} {\bibfnamefont {C.}~\bibnamefont {Sun}},\ and\
  \bibinfo {author} {\bibfnamefont {D.}~\bibnamefont {Lohse}},\ }\bibfield
  {title} {\bibinfo {title} {Multiple states in highly turbulent
  taylor--couette flow},\ }\href@noop {} {\bibfield  {journal} {\bibinfo
  {journal} {Nat. Commun.}\ }\textbf {\bibinfo {volume} {5}},\ \bibinfo {pages}
  {1} (\bibinfo {year} {2014})}\BibitemShut {NoStop}%
\bibitem [{\citenamefont {Smits}\ \emph {et~al.}(2011)\citenamefont {Smits},
  \citenamefont {McKeon},\ and\ \citenamefont {Marusic}}]{Smits2011review}%
  \BibitemOpen
  \bibfield  {author} {\bibinfo {author} {\bibfnamefont {A.~J.}\ \bibnamefont
  {Smits}}, \bibinfo {author} {\bibfnamefont {B.~J.}\ \bibnamefont {McKeon}},\
  and\ \bibinfo {author} {\bibfnamefont {I.}~\bibnamefont {Marusic}},\
  }\bibfield  {title} {\bibinfo {title} {High–{R}eynolds number wall
  turbulence},\ }\href {https://doi.org/10.1146/annurev-fluid-122109-160753}
  {\bibfield  {journal} {\bibinfo  {journal} {Annu. Rev. Fluid Mech.}\ }\textbf
  {\bibinfo {volume} {43}},\ \bibinfo {pages} {353} (\bibinfo {year}
  {2011})}\BibitemShut {NoStop}%
\bibitem [{\citenamefont {Wang}\ \emph
  {et~al.}(2019{\natexlab{a}})\citenamefont {Wang}, \citenamefont {Zhang},\
  and\ \citenamefont {Katz}}]{Katz2019pressure}%
  \BibitemOpen
  \bibfield  {author} {\bibinfo {author} {\bibfnamefont {J.}~\bibnamefont
  {Wang}}, \bibinfo {author} {\bibfnamefont {C.}~\bibnamefont {Zhang}},\ and\
  \bibinfo {author} {\bibfnamefont {J.}~\bibnamefont {Katz}},\ }\bibfield
  {title} {\bibinfo {title} {Gpu-based, parallel-line, omni-directional
  integration of measured pressure gradient field to obtain the 3{D} pressure
  distribution},\ }\href@noop {} {\bibfield  {journal} {\bibinfo  {journal}
  {Exp. Fluids}\ }\textbf {\bibinfo {volume} {60}},\ \bibinfo {pages} {1}
  (\bibinfo {year} {2019}{\natexlab{a}})}\BibitemShut {NoStop}%
\bibitem [{\citenamefont {Stuart}\ and\ \citenamefont
  {Zygalakis}(2015)}]{Stuart2015}%
  \BibitemOpen
  \bibfield  {author} {\bibinfo {author} {\bibfnamefont {A.}~\bibnamefont
  {Stuart}}\ and\ \bibinfo {author} {\bibfnamefont {K.}~\bibnamefont
  {Zygalakis}},\ }\href@noop {} {\emph {\bibinfo {title} {Data Assimilation: A
  Mathematical Introduction}}},\ \bibinfo {type} {Tech. Rep.}\ (\bibinfo
  {institution} {Oak Ridge National Lab.(ORNL), Oak Ridge, TN (United
  States)},\ \bibinfo {year} {2015})\BibitemShut {NoStop}%
\bibitem [{\citenamefont {Mons}\ \emph {et~al.}(2016)\citenamefont {Mons},
  \citenamefont {Chassaing}, \citenamefont {Gomez},\ and\ \citenamefont
  {Sagaut}}]{mons2016reconstruction}%
  \BibitemOpen
  \bibfield  {author} {\bibinfo {author} {\bibfnamefont {V.}~\bibnamefont
  {Mons}}, \bibinfo {author} {\bibfnamefont {J.-C.}\ \bibnamefont {Chassaing}},
  \bibinfo {author} {\bibfnamefont {T.}~\bibnamefont {Gomez}},\ and\ \bibinfo
  {author} {\bibfnamefont {P.}~\bibnamefont {Sagaut}},\ }\bibfield  {title}
  {\bibinfo {title} {Reconstruction of unsteady viscous flows using data
  assimilation schemes},\ }\href@noop {} {\bibfield  {journal} {\bibinfo
  {journal} {J.~Comput. Phys.}\ }\textbf {\bibinfo {volume} {316}},\ \bibinfo
  {pages} {255} (\bibinfo {year} {2016})}\BibitemShut {NoStop}%
\bibitem [{\citenamefont {Wang}\ and\ \citenamefont {Zaki}(2021)}]{mengze2021}%
  \BibitemOpen
  \bibfield  {author} {\bibinfo {author} {\bibfnamefont {M.}~\bibnamefont
  {Wang}}\ and\ \bibinfo {author} {\bibfnamefont {T.~A.}\ \bibnamefont
  {Zaki}},\ }\bibfield  {title} {\bibinfo {title} {State estimation in
  turbulent channel flow from limited observations},\ }\href@noop {} {\bibfield
   {journal} {\bibinfo  {journal} {J.~Fluid Mech.}\ }\textbf {\bibinfo {volume}
  {917}},\ \bibinfo {pages} {A9} (\bibinfo {year} {2021})}\BibitemShut
  {NoStop}%
\bibitem [{\citenamefont {Wang}\ \emph
  {et~al.}(2019{\natexlab{b}})\citenamefont {Wang}, \citenamefont {Hasegawa},\
  and\ \citenamefont {Zaki}}]{wang_hasegawa_zaki_2019}%
  \BibitemOpen
  \bibfield  {author} {\bibinfo {author} {\bibfnamefont {Q.}~\bibnamefont
  {Wang}}, \bibinfo {author} {\bibfnamefont {Y.}~\bibnamefont {Hasegawa}},\
  and\ \bibinfo {author} {\bibfnamefont {T.~A.}\ \bibnamefont {Zaki}},\
  }\bibfield  {title} {\bibinfo {title} {Spatial reconstruction of steady
  scalar sources from remote measurements in turbulent flow},\ }\href
  {https://doi.org/10.1017/jfm.2019.241} {\bibfield  {journal} {\bibinfo
  {journal} {J.~Fluid Mech.}\ }\textbf {\bibinfo {volume} {870}},\ \bibinfo
  {pages} {316–352} (\bibinfo {year} {2019}{\natexlab{b}})}\BibitemShut
  {NoStop}%
\bibitem [{\citenamefont {Mons}\ \emph {et~al.}(2019)\citenamefont {Mons},
  \citenamefont {Wang},\ and\ \citenamefont {Zaki}}]{mons2019kriging}%
  \BibitemOpen
  \bibfield  {author} {\bibinfo {author} {\bibfnamefont {V.}~\bibnamefont
  {Mons}}, \bibinfo {author} {\bibfnamefont {Q.}~\bibnamefont {Wang}},\ and\
  \bibinfo {author} {\bibfnamefont {T.~A.}\ \bibnamefont {Zaki}},\ }\bibfield
  {title} {\bibinfo {title} {Kriging-enhanced ensemble variational data
  assimilation for scalar-source identification in turbulent environments},\
  }\href@noop {} {\bibfield  {journal} {\bibinfo  {journal} {J.~Comput. Phys.}\
  }\textbf {\bibinfo {volume} {398}},\ \bibinfo {pages} {108856} (\bibinfo
  {year} {2019})}\BibitemShut {NoStop}%
\bibitem [{\citenamefont {Buchta}\ and\ \citenamefont
  {Zaki}(2021)}]{buchta2021envar}%
  \BibitemOpen
  \bibfield  {author} {\bibinfo {author} {\bibfnamefont {D.~A.}\ \bibnamefont
  {Buchta}}\ and\ \bibinfo {author} {\bibfnamefont {T.~A.}\ \bibnamefont
  {Zaki}},\ }\bibfield  {title} {\bibinfo {title} {Observation-infused
  simulations of high-speed boundary-layer transition},\ }\href
  {https://doi.org/10.1017/jfm.2021.172} {\bibfield  {journal} {\bibinfo
  {journal} {J.~Fluid Mech.}\ }\textbf {\bibinfo {volume} {916}},\ \bibinfo
  {pages} {A44} (\bibinfo {year} {2021})}\BibitemShut {NoStop}%
\bibitem [{\citenamefont {Encinar}\ and\ \citenamefont
  {Jim{\'e}nez}(2019)}]{encinar2019logarithmic}%
  \BibitemOpen
  \bibfield  {author} {\bibinfo {author} {\bibfnamefont {M.~P.}\ \bibnamefont
  {Encinar}}\ and\ \bibinfo {author} {\bibfnamefont {J.}~\bibnamefont
  {Jim{\'e}nez}},\ }\bibfield  {title} {\bibinfo {title} {Logarithmic-layer
  turbulence: A view from the wall},\ }\href@noop {} {\bibfield  {journal}
  {\bibinfo  {journal} {Phys. Rev. Fluids}\ }\textbf {\bibinfo {volume} {4}},\
  \bibinfo {pages} {114603} (\bibinfo {year} {2019})}\BibitemShut {NoStop}%
\bibitem [{\citenamefont {Chandramouli}\ \emph {et~al.}(2020)\citenamefont
  {Chandramouli}, \citenamefont {M{\'e}min},\ and\ \citenamefont
  {Heitz}}]{Chandramouli2020}%
  \BibitemOpen
  \bibfield  {author} {\bibinfo {author} {\bibfnamefont {P.}~\bibnamefont
  {Chandramouli}}, \bibinfo {author} {\bibfnamefont {E.}~\bibnamefont
  {M{\'e}min}},\ and\ \bibinfo {author} {\bibfnamefont {D.}~\bibnamefont
  {Heitz}},\ }\bibfield  {title} {\bibinfo {title} {4{D} large scale
  variational data assimilation of a turbulent flow with a dynamics error
  model},\ }\href@noop {} {\bibfield  {journal} {\bibinfo  {journal}
  {J.~Comput. Phys.}\ ,\ \bibinfo {pages} {109446}} (\bibinfo {year}
  {2020})}\BibitemShut {NoStop}%
\bibitem [{\citenamefont {Raissi}\ \emph {et~al.}(2019)\citenamefont {Raissi},
  \citenamefont {Perdikaris},\ and\ \citenamefont
  {Karniadakis}}]{Raissi2019PINN}%
  \BibitemOpen
  \bibfield  {author} {\bibinfo {author} {\bibfnamefont {M.}~\bibnamefont
  {Raissi}}, \bibinfo {author} {\bibfnamefont {P.}~\bibnamefont {Perdikaris}},\
  and\ \bibinfo {author} {\bibfnamefont {G.~E.}\ \bibnamefont {Karniadakis}},\
  }\bibfield  {title} {\bibinfo {title} {Physics-informed neural networks: A
  deep learning framework for solving forward and inverse problems involving
  nonlinear partial differential equations},\ }\href@noop {} {\bibfield
  {journal} {\bibinfo  {journal} {J.~Comput. Phys.}\ }\textbf {\bibinfo
  {volume} {378}},\ \bibinfo {pages} {686} (\bibinfo {year}
  {2019})}\BibitemShut {NoStop}%
\bibitem [{\citenamefont {Du}\ and\ \citenamefont {Zaki}(2021)}]{Du2021EDNN}%
  \BibitemOpen
  \bibfield  {author} {\bibinfo {author} {\bibfnamefont {Y.}~\bibnamefont
  {Du}}\ and\ \bibinfo {author} {\bibfnamefont {T.}~\bibnamefont {Zaki}},\
  }\bibfield  {title} {\bibinfo {title} {Evolutional deep neural network},\
  }\href@noop {} {\bibfield  {journal} {\bibinfo  {journal} {arXiv preprint
  arXiv:2103.09959}\ } (\bibinfo {year} {2021})}\BibitemShut {NoStop}%
\bibitem [{\citenamefont {Evensen}(1994)}]{Evensen1994EnKF}%
  \BibitemOpen
  \bibfield  {author} {\bibinfo {author} {\bibfnamefont {G.}~\bibnamefont
  {Evensen}},\ }\bibfield  {title} {\bibinfo {title} {Sequential data
  assimilation with a nonlinear quasi-geostrophic model using monte carlo
  methods to forecast error statistics},\ }\href@noop {} {\bibfield  {journal}
  {\bibinfo  {journal} {J. Geophys. Res. Oceans}\ }\textbf {\bibinfo {volume}
  {99}},\ \bibinfo {pages} {10143} (\bibinfo {year} {1994})}\BibitemShut
  {NoStop}%
\bibitem [{\citenamefont {Suzuki}(2012)}]{Suzuki2012}%
  \BibitemOpen
  \bibfield  {author} {\bibinfo {author} {\bibfnamefont {T.}~\bibnamefont
  {Suzuki}},\ }\bibfield  {title} {\bibinfo {title} {Reduced-order
  {K}alman-filtered hybrid simulation combining particle tracking velocimetry
  and direct numerical simulation},\ }\href@noop {} {\bibfield  {journal}
  {\bibinfo  {journal} {J.~Fluid Mech.}\ }\textbf {\bibinfo {volume} {709}},\
  \bibinfo {pages} {249–288} (\bibinfo {year} {2012})}\BibitemShut {NoStop}%
\bibitem [{\citenamefont {Dimet}\ and\ \citenamefont
  {Talagrand}(1986)}]{Dimet1986_4dvar}%
  \BibitemOpen
  \bibfield  {author} {\bibinfo {author} {\bibfnamefont {F.-X.~L.}\
  \bibnamefont {Dimet}}\ and\ \bibinfo {author} {\bibfnamefont
  {O.}~\bibnamefont {Talagrand}},\ }\bibfield  {title} {\bibinfo {title}
  {Variational algorithms for analysis and assimilation of meteorological
  observations: theoretical aspects},\ }\href@noop {} {\bibfield  {journal}
  {\bibinfo  {journal} {Tellus A: Dyn. Meteorol. Oceanogr.}\ }\textbf {\bibinfo
  {volume} {38}},\ \bibinfo {pages} {97} (\bibinfo {year} {1986})}\BibitemShut
  {NoStop}%
\bibitem [{\citenamefont {Wang}\ \emph
  {et~al.}(2019{\natexlab{c}})\citenamefont {Wang}, \citenamefont {Wang},\ and\
  \citenamefont {Zaki}}]{mengze2019discrete}%
  \BibitemOpen
  \bibfield  {author} {\bibinfo {author} {\bibfnamefont {M.}~\bibnamefont
  {Wang}}, \bibinfo {author} {\bibfnamefont {Q.}~\bibnamefont {Wang}},\ and\
  \bibinfo {author} {\bibfnamefont {T.~A.}\ \bibnamefont {Zaki}},\ }\bibfield
  {title} {\bibinfo {title} {Discrete adjoint of fractional-step incompressible
  {N}avier-{S}tokes solver in curvilinear coordinates and application to data
  assimilation},\ }\bibfield  {journal} {\bibinfo  {journal} {J.~Comput.
  Phys.}\ }\href {https://doi.org/https://doi.org/10.1016/j.jcp.2019.06.065}
  {https://doi.org/10.1016/j.jcp.2019.06.065} (\bibinfo {year}
  {2019}{\natexlab{c}})\BibitemShut {NoStop}%
\bibitem [{\citenamefont {Chandramoorthy}\ \emph {et~al.}(2019)\citenamefont
  {Chandramoorthy}, \citenamefont {Fernandez}, \citenamefont {Talnikar},\ and\
  \citenamefont {Wang}}]{chandramoorthy2019feasibility}%
  \BibitemOpen
  \bibfield  {author} {\bibinfo {author} {\bibfnamefont {N.}~\bibnamefont
  {Chandramoorthy}}, \bibinfo {author} {\bibfnamefont {P.}~\bibnamefont
  {Fernandez}}, \bibinfo {author} {\bibfnamefont {C.}~\bibnamefont
  {Talnikar}},\ and\ \bibinfo {author} {\bibfnamefont {Q.}~\bibnamefont
  {Wang}},\ }\bibfield  {title} {\bibinfo {title} {Feasibility analysis of
  ensemble sensitivity computation in turbulent flows},\ }\href@noop {}
  {\bibfield  {journal} {\bibinfo  {journal} {AIAA J.}\ }\textbf {\bibinfo
  {volume} {57}},\ \bibinfo {pages} {4514} (\bibinfo {year}
  {2019})}\BibitemShut {NoStop}%
\bibitem [{\citenamefont {Eyink}\ \emph {et~al.}(2004)\citenamefont {Eyink},
  \citenamefont {Haine},\ and\ \citenamefont {Lea}}]{eyink2004ruelle}%
  \BibitemOpen
  \bibfield  {author} {\bibinfo {author} {\bibfnamefont {G.}~\bibnamefont
  {Eyink}}, \bibinfo {author} {\bibfnamefont {T.}~\bibnamefont {Haine}},\ and\
  \bibinfo {author} {\bibfnamefont {D.}~\bibnamefont {Lea}},\ }\bibfield
  {title} {\bibinfo {title} {Ruelle's linear response formula, ensemble adjoint
  schemes and {L}{\'e}vy flights},\ }\href@noop {} {\bibfield  {journal}
  {\bibinfo  {journal} {Nonlinearity}\ }\textbf {\bibinfo {volume} {17}},\
  \bibinfo {pages} {1867} (\bibinfo {year} {2004})}\BibitemShut {NoStop}%
\bibitem [{\citenamefont {Fisher}\ and\ \citenamefont
  {Auvinen}(2012)}]{Fisher2012}%
  \BibitemOpen
  \bibfield  {author} {\bibinfo {author} {\bibfnamefont {M.}~\bibnamefont
  {Fisher}}\ and\ \bibinfo {author} {\bibfnamefont {H.}~\bibnamefont
  {Auvinen}},\ }\bibfield  {title} {\bibinfo {title} {Long window 4d-var},\
  }in\ \href {https://www.ecmwf.int/node/9410} {\emph {\bibinfo {booktitle}
  {Seminar on Data assimilation for atmosphere and ocean, 6-9 September
  2011}}},\ \bibinfo {organization} {ECMWF}\ (\bibinfo  {publisher} {ECMWF},\
  \bibinfo {address} {Shinfield Park, Reading},\ \bibinfo {year} {2012})\ pp.\
  \bibinfo {pages} {189--202}\BibitemShut {NoStop}%
\bibitem [{\citenamefont {Nocedal}(1980)}]{LBFGS}%
  \BibitemOpen
  \bibfield  {author} {\bibinfo {author} {\bibfnamefont {J.}~\bibnamefont
  {Nocedal}},\ }\bibfield  {title} {\bibinfo {title} {Updating {quasi-Newton}
  matrices with limited storage},\ }\href@noop {} {\bibfield  {journal}
  {\bibinfo  {journal} {Math. Comput.}\ }\textbf {\bibinfo {volume} {35}},\
  \bibinfo {pages} {773} (\bibinfo {year} {1980})}\BibitemShut {NoStop}%
\bibitem [{\citenamefont {Zaki}(2013)}]{zaki2013streaks}%
  \BibitemOpen
  \bibfield  {author} {\bibinfo {author} {\bibfnamefont {T.~A.}\ \bibnamefont
  {Zaki}},\ }\bibfield  {title} {\bibinfo {title} {From streaks to spots and on
  to turbulence: exploring the dynamics of boundary layer transition},\
  }\href@noop {} {\bibfield  {journal} {\bibinfo  {journal} {Flow Turbul.
  Combust.}\ }\textbf {\bibinfo {volume} {91}},\ \bibinfo {pages} {451}
  (\bibinfo {year} {2013})}\BibitemShut {NoStop}%
\bibitem [{\citenamefont {Marxen}\ and\ \citenamefont
  {Zaki}(2019)}]{marxen2019turbulence}%
  \BibitemOpen
  \bibfield  {author} {\bibinfo {author} {\bibfnamefont {O.}~\bibnamefont
  {Marxen}}\ and\ \bibinfo {author} {\bibfnamefont {T.~A.}\ \bibnamefont
  {Zaki}},\ }\bibfield  {title} {\bibinfo {title} {Turbulence in intermittent
  transitional boundary layers and in turbulence spots},\ }\href@noop {}
  {\bibfield  {journal} {\bibinfo  {journal} {J.~Fluid Mech.}\ }\textbf
  {\bibinfo {volume} {860}},\ \bibinfo {pages} {350} (\bibinfo {year}
  {2019})}\BibitemShut {NoStop}%
\bibitem [{\citenamefont {Wu}\ \emph {et~al.}(2020)\citenamefont {Wu},
  \citenamefont {Zaki},\ and\ \citenamefont {Meneveau}}]{wupnas2020}%
  \BibitemOpen
  \bibfield  {author} {\bibinfo {author} {\bibfnamefont {Z.}~\bibnamefont
  {Wu}}, \bibinfo {author} {\bibfnamefont {T.~A.}\ \bibnamefont {Zaki}},\ and\
  \bibinfo {author} {\bibfnamefont {C.}~\bibnamefont {Meneveau}},\ }\bibfield
  {title} {\bibinfo {title} {High-{R}eynolds-number fractal signature of
  nascent turbulence during transition},\ }\href@noop {} {\bibfield  {journal}
  {\bibinfo  {journal} {Proc. Natl. Acad. Sci. U.S.A.}\ }\textbf {\bibinfo
  {volume} {117}},\ \bibinfo {pages} {3461} (\bibinfo {year}
  {2020})}\BibitemShut {NoStop}%
\bibitem [{\citenamefont {Reynolds}(1883)}]{Reynolds1883}%
  \BibitemOpen
  \bibfield  {author} {\bibinfo {author} {\bibfnamefont {O.}~\bibnamefont
  {Reynolds}},\ }\bibfield  {title} {\bibinfo {title} {{XXIX}. {A}n
  experimental investigation of the circumstances which determine whether the
  motion of water shall be direct or sinuous, and of the law of resistance in
  parallel channels},\ }\href@noop {} {\bibfield  {journal} {\bibinfo
  {journal} {Phil. Trans. R. Soc.}\ ,\ \bibinfo {pages} {935}} (\bibinfo {year}
  {1883})}\BibitemShut {NoStop}%
\bibitem [{\citenamefont {Avila}\ \emph {et~al.}(2011)\citenamefont {Avila},
  \citenamefont {Moxey}, \citenamefont {de~Lozar}, \citenamefont {Avila},
  \citenamefont {Barkley},\ and\ \citenamefont {Hof}}]{Avila2011pipe}%
  \BibitemOpen
  \bibfield  {author} {\bibinfo {author} {\bibfnamefont {K.}~\bibnamefont
  {Avila}}, \bibinfo {author} {\bibfnamefont {D.}~\bibnamefont {Moxey}},
  \bibinfo {author} {\bibfnamefont {A.}~\bibnamefont {de~Lozar}}, \bibinfo
  {author} {\bibfnamefont {M.}~\bibnamefont {Avila}}, \bibinfo {author}
  {\bibfnamefont {D.}~\bibnamefont {Barkley}},\ and\ \bibinfo {author}
  {\bibfnamefont {B.}~\bibnamefont {Hof}},\ }\bibfield  {title} {\bibinfo
  {title} {The onset of turbulence in pipe flow},\ }\href@noop {} {\bibfield
  {journal} {\bibinfo  {journal} {Science}\ }\textbf {\bibinfo {volume}
  {333}},\ \bibinfo {pages} {192} (\bibinfo {year} {2011})}\BibitemShut
  {NoStop}%
\bibitem [{\citenamefont {Lorenz}(1963)}]{Lorenz1963}%
  \BibitemOpen
  \bibfield  {author} {\bibinfo {author} {\bibfnamefont {E.~N.}\ \bibnamefont
  {Lorenz}},\ }\bibfield  {title} {\bibinfo {title} {Deterministic nonperiodic
  flow},\ }\href@noop {} {\bibfield  {journal} {\bibinfo  {journal} {J.~Atmos.
  Sci.}\ }\textbf {\bibinfo {volume} {20}},\ \bibinfo {pages} {130} (\bibinfo
  {year} {1963})}\BibitemShut {NoStop}%
\bibitem [{\citenamefont {Saltzman}(1962)}]{Saltzman1962}%
  \BibitemOpen
  \bibfield  {author} {\bibinfo {author} {\bibfnamefont {B.}~\bibnamefont
  {Saltzman}},\ }\bibfield  {title} {\bibinfo {title} {Finite amplitude free
  convection as an initial value problem-{I}},\ }\href@noop {} {\bibfield
  {journal} {\bibinfo  {journal} {J.~Atmos. Sci.}\ }\textbf {\bibinfo {volume}
  {19}},\ \bibinfo {pages} {329} (\bibinfo {year} {1962})}\BibitemShut
  {NoStop}%
\bibitem [{\citenamefont {Nikitin}(2018)}]{nikitin2018characteristics}%
  \BibitemOpen
  \bibfield  {author} {\bibinfo {author} {\bibfnamefont {N.}~\bibnamefont
  {Nikitin}},\ }\bibfield  {title} {\bibinfo {title} {Characteristics of the
  leading {L}yapunov vector in a turbulent channel flow},\ }\href@noop {}
  {\bibfield  {journal} {\bibinfo  {journal} {J.~Fluid Mech.}\ }\textbf
  {\bibinfo {volume} {849}},\ \bibinfo {pages} {942} (\bibinfo {year}
  {2018})}\BibitemShut {NoStop}%
\bibitem [{\citenamefont {Abrahamson}\ and\ \citenamefont
  {Lonnes}(1995)}]{Abrahamson1995}%
  \BibitemOpen
  \bibfield  {author} {\bibinfo {author} {\bibfnamefont {S.}~\bibnamefont
  {Abrahamson}}\ and\ \bibinfo {author} {\bibfnamefont {S.}~\bibnamefont
  {Lonnes}},\ }\bibfield  {title} {\bibinfo {title} {Uncertainty in calculating
  vorticity from {2D} velocity fields using circulation and least-squares
  approaches},\ }\href@noop {} {\bibfield  {journal} {\bibinfo  {journal} {Exp.
  Fluids}\ }\textbf {\bibinfo {volume} {20}},\ \bibinfo {pages} {10} (\bibinfo
  {year} {1995})}\BibitemShut {NoStop}%
\bibitem [{\citenamefont {Eyink}\ \emph
  {et~al.}(2020{\natexlab{a}})\citenamefont {Eyink}, \citenamefont {Gupta},\
  and\ \citenamefont {Zaki}}]{Eyink2020_theory}%
  \BibitemOpen
  \bibfield  {author} {\bibinfo {author} {\bibfnamefont {G.~L.}\ \bibnamefont
  {Eyink}}, \bibinfo {author} {\bibfnamefont {A.}~\bibnamefont {Gupta}},\ and\
  \bibinfo {author} {\bibfnamefont {T.~A.}\ \bibnamefont {Zaki}},\ }\bibfield
  {title} {\bibinfo {title} {Stochastic {L}agrangian dynamics of vorticity.
  {P}art 1. {G}eneral theory for viscous, incompressible fluids},\ }\href
  {https://doi.org/10.1017/jfm.2020.491} {\bibfield  {journal} {\bibinfo
  {journal} {J.~Fluid Mech.}\ }\textbf {\bibinfo {volume} {901}},\ \bibinfo
  {pages} {A2} (\bibinfo {year} {2020}{\natexlab{a}})}\BibitemShut {NoStop}%
\bibitem [{\citenamefont {Eyink}\ \emph
  {et~al.}(2020{\natexlab{b}})\citenamefont {Eyink}, \citenamefont {Gupta},\
  and\ \citenamefont {Zaki}}]{Eyink2020_channel}%
  \BibitemOpen
  \bibfield  {author} {\bibinfo {author} {\bibfnamefont {G.~L.}\ \bibnamefont
  {Eyink}}, \bibinfo {author} {\bibfnamefont {A.}~\bibnamefont {Gupta}},\ and\
  \bibinfo {author} {\bibfnamefont {T.~A.}\ \bibnamefont {Zaki}},\ }\bibfield
  {title} {\bibinfo {title} {Stochastic {L}agrangian dynamics of vorticity.
  {P}art 2. {A}pplication to near-wall channel-flow turbulence},\ }\href
  {https://doi.org/10.1017/jfm.2020.492} {\bibfield  {journal} {\bibinfo
  {journal} {J.~Fluid Mech.}\ }\textbf {\bibinfo {volume} {901}},\ \bibinfo
  {pages} {A3} (\bibinfo {year} {2020}{\natexlab{b}})}\BibitemShut {NoStop}%
\bibitem [{\citenamefont {Bewley}\ and\ \citenamefont
  {Protas}(2004)}]{bewley2004skin}%
  \BibitemOpen
  \bibfield  {author} {\bibinfo {author} {\bibfnamefont {T.~R.}\ \bibnamefont
  {Bewley}}\ and\ \bibinfo {author} {\bibfnamefont {B.}~\bibnamefont
  {Protas}},\ }\bibfield  {title} {\bibinfo {title} {Skin friction and
  pressure: the “footprints” of turbulence},\ }\href@noop {} {\bibfield
  {journal} {\bibinfo  {journal} {Phys. D: Nonlinear Phenom.}\ }\textbf
  {\bibinfo {volume} {196}},\ \bibinfo {pages} {28} (\bibinfo {year}
  {2004})}\BibitemShut {NoStop}%
\bibitem [{\citenamefont {Suzuki}\ and\ \citenamefont
  {Hasegawa}(2017)}]{hasegawa2016estimation}%
  \BibitemOpen
  \bibfield  {author} {\bibinfo {author} {\bibfnamefont {T.}~\bibnamefont
  {Suzuki}}\ and\ \bibinfo {author} {\bibfnamefont {Y.}~\bibnamefont
  {Hasegawa}},\ }\bibfield  {title} {\bibinfo {title} {Estimation of turbulent
  channel flow at {$Re_{\tau}=100$} based on the wall measurement using a
  simple sequential approach},\ }\href@noop {} {\bibfield  {journal} {\bibinfo
  {journal} {J.~Fluid Mech.}\ }\textbf {\bibinfo {volume} {830}},\ \bibinfo
  {pages} {760} (\bibinfo {year} {2017})}\BibitemShut {NoStop}%
\bibitem [{\citenamefont {Rowley}(2005)}]{Rowley2005gramian}%
  \BibitemOpen
  \bibfield  {author} {\bibinfo {author} {\bibfnamefont {C.~W.}\ \bibnamefont
  {Rowley}},\ }\bibfield  {title} {\bibinfo {title} {Model reduction for
  fluids, using balanced proper orthogonal decomposition},\ }\href@noop {}
  {\bibfield  {journal} {\bibinfo  {journal} {Int. J. Bifurcat. Chaos}\
  }\textbf {\bibinfo {volume} {15}},\ \bibinfo {pages} {997} (\bibinfo {year}
  {2005})}\BibitemShut {NoStop}%
\bibitem [{\citenamefont {Wang}\ \emph {et~al.}(2021)\citenamefont {Wang},
  \citenamefont {Wang},\ and\ \citenamefont {Zaki}}]{Qi2021}%
  \BibitemOpen
  \bibfield  {author} {\bibinfo {author} {\bibfnamefont {Q.}~\bibnamefont
  {Wang}}, \bibinfo {author} {\bibfnamefont {M.}~\bibnamefont {Wang}},\ and\
  \bibinfo {author} {\bibfnamefont {T.}~\bibnamefont {Zaki}},\ }\bibfield
  {title} {\bibinfo {title} {From wall observations to turbulence: {T}he
  difficulty of flow reconstruction},\ }\href@noop {} {\bibfield  {journal}
  {\bibinfo  {journal} {arXiv preprint arXiv:2106.09169}\ } (\bibinfo {year}
  {2021})}\BibitemShut {NoStop}%
\bibitem [{\citenamefont {Hunt}\ and\ \citenamefont
  {Durbin}(1999)}]{hunt1999perturbed}%
  \BibitemOpen
  \bibfield  {author} {\bibinfo {author} {\bibfnamefont {J.}~\bibnamefont
  {Hunt}}\ and\ \bibinfo {author} {\bibfnamefont {P.}~\bibnamefont {Durbin}},\
  }\bibfield  {title} {\bibinfo {title} {Perturbed vortical layers and shear
  sheltering},\ }\href@noop {} {\bibfield  {journal} {\bibinfo  {journal}
  {Fluid Dyn. Res.}\ }\textbf {\bibinfo {volume} {24}},\ \bibinfo {pages} {375}
  (\bibinfo {year} {1999})}\BibitemShut {NoStop}%
\bibitem [{\citenamefont {Zaki}\ and\ \citenamefont
  {Saha}(2009)}]{zaki2009shear}%
  \BibitemOpen
  \bibfield  {author} {\bibinfo {author} {\bibfnamefont {T.~A.}\ \bibnamefont
  {Zaki}}\ and\ \bibinfo {author} {\bibfnamefont {S.}~\bibnamefont {Saha}},\
  }\bibfield  {title} {\bibinfo {title} {On shear sheltering and the structure
  of vortical modes in single-and two-fluid boundary layers},\ }\href@noop {}
  {\bibfield  {journal} {\bibinfo  {journal} {J.~Fluid Mech.}\ }\textbf
  {\bibinfo {volume} {626}},\ \bibinfo {pages} {111} (\bibinfo {year}
  {2009})}\BibitemShut {NoStop}%
\bibitem [{\citenamefont {Mathis}\ \emph {et~al.}(2009)\citenamefont {Mathis},
  \citenamefont {Hutchins},\ and\ \citenamefont {Marusic}}]{mathis2009large}%
  \BibitemOpen
  \bibfield  {author} {\bibinfo {author} {\bibfnamefont {R.}~\bibnamefont
  {Mathis}}, \bibinfo {author} {\bibfnamefont {N.}~\bibnamefont {Hutchins}},\
  and\ \bibinfo {author} {\bibfnamefont {I.}~\bibnamefont {Marusic}},\
  }\bibfield  {title} {\bibinfo {title} {Large-scale amplitude modulation of
  the small-scale structures in turbulent boundary layers},\ }\href@noop {}
  {\bibfield  {journal} {\bibinfo  {journal} {J.~Fluid Mech.}\ } (\bibinfo
  {year} {2009})}\BibitemShut {NoStop}%
\bibitem [{\citenamefont {Abe}\ \emph {et~al.}(2004)\citenamefont {Abe},
  \citenamefont {Kawamura},\ and\ \citenamefont {Choi}}]{abe2004very}%
  \BibitemOpen
  \bibfield  {author} {\bibinfo {author} {\bibfnamefont {H.}~\bibnamefont
  {Abe}}, \bibinfo {author} {\bibfnamefont {H.}~\bibnamefont {Kawamura}},\ and\
  \bibinfo {author} {\bibfnamefont {H.}~\bibnamefont {Choi}},\ }\bibfield
  {title} {\bibinfo {title} {Very large-scale structures and their effects on
  the wall shear-stress fluctuations in a turbulent channel flow up to re
  $\tau$= 640},\ }\href@noop {} {\bibfield  {journal} {\bibinfo  {journal} {J.
  Fluids Eng.}\ }\textbf {\bibinfo {volume} {126}},\ \bibinfo {pages} {835}
  (\bibinfo {year} {2004})}\BibitemShut {NoStop}%
\bibitem [{\citenamefont {Hwang}\ \emph {et~al.}(2016)\citenamefont {Hwang},
  \citenamefont {Lee}, \citenamefont {Sung},\ and\ \citenamefont
  {Zaki}}]{hwang2016inner}%
  \BibitemOpen
  \bibfield  {author} {\bibinfo {author} {\bibfnamefont {J.}~\bibnamefont
  {Hwang}}, \bibinfo {author} {\bibfnamefont {J.}~\bibnamefont {Lee}}, \bibinfo
  {author} {\bibfnamefont {H.~J.}\ \bibnamefont {Sung}},\ and\ \bibinfo
  {author} {\bibfnamefont {T.~A.}\ \bibnamefont {Zaki}},\ }\bibfield  {title}
  {\bibinfo {title} {Inner-outer interactions of large-scale structures in
  turbulent channel flow},\ }\href@noop {} {\bibfield  {journal} {\bibinfo
  {journal} {J.~Fluid Mech.}\ }\textbf {\bibinfo {volume} {790}},\ \bibinfo
  {pages} {128} (\bibinfo {year} {2016})}\BibitemShut {NoStop}%
\bibitem [{\citenamefont {You}\ and\ \citenamefont {Zaki}(2019)}]{you2019tbl}%
  \BibitemOpen
  \bibfield  {author} {\bibinfo {author} {\bibfnamefont {J.}~\bibnamefont
  {You}}\ and\ \bibinfo {author} {\bibfnamefont {T.~A.}\ \bibnamefont {Zaki}},\
  }\bibfield  {title} {\bibinfo {title} {Conditional statistics and flow
  structures in turbulent boundary layers buffeted by free-stream
  disturbances},\ }\href {https://doi.org/10.1017/jfm.2019.104} {\bibfield
  {journal} {\bibinfo  {journal} {J.~Fluid Mech.}\ }\textbf {\bibinfo {volume}
  {866}},\ \bibinfo {pages} {526–566} (\bibinfo {year} {2019})}\BibitemShut
  {NoStop}%
\bibitem [{\citenamefont {Li}\ \emph {et~al.}(2020)\citenamefont {Li},
  \citenamefont {Zhang}, \citenamefont {Dong},\ and\ \citenamefont
  {Abdullah}}]{li2020}%
  \BibitemOpen
  \bibfield  {author} {\bibinfo {author} {\bibfnamefont {Y.}~\bibnamefont
  {Li}}, \bibinfo {author} {\bibfnamefont {J.}~\bibnamefont {Zhang}}, \bibinfo
  {author} {\bibfnamefont {G.}~\bibnamefont {Dong}},\ and\ \bibinfo {author}
  {\bibfnamefont {N.~S.}\ \bibnamefont {Abdullah}},\ }\bibfield  {title}
  {\bibinfo {title} {Small-scale reconstruction in three-dimensional kolmogorov
  flows using four-dimensional variational data assimilation},\ }\href@noop {}
  {\bibfield  {journal} {\bibinfo  {journal} {J.~Fluid Mech.}\ }\textbf
  {\bibinfo {volume} {885}},\ \bibinfo {pages} {A9} (\bibinfo {year}
  {2020})}\BibitemShut {NoStop}%
\bibitem [{\citenamefont {Di~Leoni}\ \emph {et~al.}(2021)\citenamefont
  {Di~Leoni}, \citenamefont {Lu}, \citenamefont {Meneveau}, \citenamefont
  {Karniadakis},\ and\ \citenamefont {Zaki}}]{PCDL2021DeepONet}%
  \BibitemOpen
  \bibfield  {author} {\bibinfo {author} {\bibfnamefont {P.~C.}\ \bibnamefont
  {Di~Leoni}}, \bibinfo {author} {\bibfnamefont {L.}~\bibnamefont {Lu}},
  \bibinfo {author} {\bibfnamefont {C.}~\bibnamefont {Meneveau}}, \bibinfo
  {author} {\bibfnamefont {G.}~\bibnamefont {Karniadakis}},\ and\ \bibinfo
  {author} {\bibfnamefont {T.~A.}\ \bibnamefont {Zaki}},\ }\bibfield  {title}
  {\bibinfo {title} {Deeponet prediction of linear instability waves in
  high-speed boundary layers},\ }\href@noop {} {\bibfield  {journal} {\bibinfo
  {journal} {arXiv preprint arXiv:2105.08697}\ } (\bibinfo {year}
  {2021})}\BibitemShut {NoStop}%
\end{thebibliography}%

\end{document}